\documentclass[fleqn,usenatbib]{mnras}

\usepackage{newtxtext,newtxmath}

\usepackage[T1]{fontenc}

\DeclareRobustCommand{\VAN}[3]{#2}
\let\VANthebibliography\thebibliography
\def\thebibliography{\DeclareRobustCommand{\VAN}[3]{##3}\VANthebibliography}

\usepackage{graphicx}	
\usepackage{amsmath}	

\usepackage{xcolor}     

\title[Nonlinear theory of resonant warps]{Nonlinear resonant torus oscillations as a model of Keplerian disc warp dynamics}

\author[Fairbairn \& Ogilvie]{
Callum W. Fairbairn$^{1}$\thanks{E-mail: cwf29@cam.ac.uk}
and Gordon I. Ogilvie,$^{1}$\thanks{E-mail: gio10@cam.ac.uk}
\\
$^{1}$Department of Applied Mathematics and Theoretical Physics, University of Cambridge, Centre for Mathematical Sciences,\\
Wilberforce Road, Cambridge CB3 0WA, UK
}

\date{Accepted XXX. Received YYY; in original form ZZZ}

\pubyear{2021}

\begin{document}
\label{firstpage}
\pagerange{\pageref{firstpage}--\pageref{lastpage}}
\maketitle

\begin{abstract}
Observations of distorted discs have highlighted the ubiquity of warps in a variety of astrophysical contexts. This has been complemented by theoretical efforts to understand the dynamics of warp evolution. Despite significant efforts to understand the dynamics of warped discs, previous work fails to address arguably the most prevalent regime -- nonlinear warps in Keplerian discs for which there is a resonance between the orbital, epicyclic and vertical oscillation frequencies. In this work, we implement a novel nonlinear ring model, developed recently by Fairbairn and Ogilvie, as a framework for understanding such resonant warp dynamics. Here we uncover two distinct nonlinear regimes as the warp amplitude is increased. Initially we find a smooth modulation theory which describes warp evolution in terms of the averaged Lagrangian of the oscillatory vertical motions of the disc. This hints towards the possibility of connecting previous warp theory under a generalised secular framework. Upon the warp amplitude exceeding a critical value, which scales as the square root of the aspect-ratio of our ring, the disc enters into a bouncing regime with extreme vertical compressions twice per orbit. We develop an impulsive theory which predicts special retrograde and prograde precessing warped solutions, which are identified numerically using our full equation set. Such solutions emphasise the essential activation of nonlinear vertical oscillations within the disc and may have important implications for energy and warp dissipation. Future work should search for this behaviour in detailed numerical studies of the internal flow structure of warped discs.
\end{abstract}




\section{Introduction}
\label{section:introduction}
\subsection{Astrophysical motivation}
\label{subsection:intro:astrophysical_motivation}
The traditional model for astrophysical discs assumes the simplest coplanar configuration with fluid streamlines on circular orbits. However, there has been growing interest in the behaviour of these systems when they become distorted by a warp. This introduces a radial variation in the inclination of the circular streamlines which might drastically alter the disc dynamics. Indeed, there is an ever expanding host of observational evidence for warped discs in a variety of contexts, which demands an improved theoretical understanding. 

Warped discs have been indirectly inferred from the long period luminosity variations of `superorbital' X-ray binary systems where a precessing warped structure periodically obscures light from a central source \citep[e.g.][]{Katz1973, Kotze2012}. In a similar vein, intensity deficits in the outer regions of protoplanetary discs may be explained by shadows cast by an inner tilted precessing disc \citep[e.g.][]{Debes2017, Muro-Arena_2020}. Comparison of radiative models with observed shadows have suggested even more extreme inclination variations in transition discs, where large radial gaps divide the inner and outer regions \citep[e.g.][]{Marino2015, Pinilla2015, Stolker2016, Benisty2017,Casassus2018}. In some systems these distinct rings are thought to form by disc tearing and breaking, as found in several numerical simulations. \cite{Nixon2012} find that Lense-Thirring torque around a spinning black hole can induce disc breaking whilst \cite{Facchini2013} find breaking of a circumbinary disc when it is sufficiently tilted with respect to the plane of the binary. Radiative post-processing of such structures produces images capable of explaining observed precessing shadows \citep{Facchini2018}. More recently, there has been an observation of the spectacular triple star system GW Orionis wherein gravitational effects may have torn the disc into independently precessing rings \citep{Kraus2020}.

These indirect cases have been complemented by direct observations of maser emission lines tracing warped galactic midplanes, as for the spiral galaxy NGC 4258 (M106) \citep{Miyoshi1995}. More recently, the \textit{Atacama Large Millimeter/submillimeter Array} (ALMA) has measured dust emission in young protostellar discs with misaligned inner and outer regions \citep[][]{Sakai2019}. ALMA has also traced gas kinematics through CO and HCO$^{+}$ molecular line emission which is consistent with warped inner regions \citep{Rosenfeld2012,Loomis2017}.

\subsection{Warped disc theory}
In order to understand this host of observational phenomena, we require theoretical models for the evolution of warped discs. Much of the mathematical language underpinning these was laid down by the work of \cite{Petterson1977a,Petterson1977b} and \cite{Hatchett1981} wherein the warp is described as a series of nested, interacting rings. Understanding the evolution is then a question of determining the time dependence of the inclination of each ring. \cite{Petterson1977a} included a viscous torque between the rings which naturally led to the diffusion of warp on a viscous timescale. However, \cite{PapaloizouPringle1983} showed that this simple model neglected the internal flow dynamics established by the warp itself, which enhance the angular momentum transport and accelerate the warp evolution. They found that the evolution is diffusive (but faster than the viscous timescale) when $\alpha > H/R$, where $\alpha$ is the Shakura-Sunyaev viscosity parameter and $H/R$ is the angular semi-thickness of the disc. Later, \cite{PapaloizouLin1995} and \cite{Lubow2000} investigated the nearly inviscid regime for which $\alpha < H/R$. Here the linearised evolution takes the form of a non-dispersive bending wave in Keplerian discs and a dispersive bending wave when the degeneracy between the epicylic and vertical frequencies is broken. 

All these models focus on linear warps, but of course it is crucial to extend this understanding into the nonlinear regime where there are observational consequences. \cite{Ogilvie1999} improved on the efforts of \cite{Pringle1992} and developed a self consistent, fully nonlinear model of diffusion in Keplerian discs and bending waves in non-Keplerian discs. This theory has been shown to agree well with numerical simulations of warps \citep{Lodato2010}. Despite such success, this model is unable to describe arguably the most important case -- inviscid Keplerian discs where the epicyclic motion is resonantly driven by the warping geometry. \cite{Ogilvie2006} attempted to explore this missing regime by performing a weakly non-linear analysis of Keplerian bending waves. However, the strongly nonlinear case still lacks a complete theory and requires further attention.

\subsection{Outline of this paper}
In order to address this problem we previously introduced a novel ring model, capable of describing the fully nonlinear hydrodynamic oscillations of an ideal, non-self gravitating torus (\citealt{FairbairnOgilvie2021}, hereafter Paper I). We found that small amplitude tilting oscillations in this local model could be identified with global linear bending waves. Indeed, our shearing box formulation effectively captures the evolution of a warp as we zoom in on a localised patch of the disc. In this picture, ring oscillations over the fast orbital timescale track the azimuthal changes in the disc geometry as the shearing box moves around the orbit. Thus tilting motions are associated with streamlines on inclined orbits and hence warps about the midplane. In this paper we aim to advance this theory and examine the nonlinear extension of bending modes. We begin by summarising the derivation and interpretation of the ring model equations in section \ref{section:ring_model}. We then motivate our analytical progress by performing some numerical runs in section \ref{section:numerical_motivation}. We will find that as the initialised warp amplitude is increased, two distinct nonlinear regimes arise. We will tackle the first in section \ref{section:smooth_modulation} by using an averaged Lagrangian method which describes a smooth modulation of the warp amplitude and phase. This behaviour drastically changes beyond some critical warp amplitude, at which point the disc enters into an extreme bouncing regime. To this end we develop an impulsive bouncing theory in sections \ref{section:bouncing_regime} and \ref{section:resonant_centres} which predicts a family of highly compressive, warped solutions. These analytical predictions are confirmed within the full ring model equation set in section \ref{section:shooting_method} before we discuss the implications for astrophysical discs and warp theory in section \ref{section:discussion}.
\section{Summary of ring model}
\label{section:ring_model}
\subsection{Model assumptions}
In Paper I we constructed a ring model for oscillating tori which will prove a useful framework in our current study. In this section we will briefly revisit the key assumptions and the resulting equations. Following the standard shearing box construction \citep[e.g.][]{Hill1878,Hawley1995}, we expand the ideal hydrodynamic equations about a local circular reference orbit at $r_0$ with angular velocity $\bmath{\Omega_0} = \Omega(r_0)\hat{\bmath{z}}$, assuming an axisymmetric potential $\Phi(r,z)$. This orbit has an attached, co-rotating coordinate system $(x, y, z)$ which is defined by $x = (r-r_0)$, $y = r_0(\phi-\Omega_0 t)$ and $z = z$, such that $x$, $y$ and $z$ are the radial, azimuthal and vertical directions respectively. This leads to the usual shearing box equations
\begin{equation}
    D\bmath{u} + 2\bmath{\Omega_0}\times\bmath{u} = -\nabla \Phi_t -\frac{1}{\rho}\nabla p,
\end{equation}
where
\begin{equation}
    D = \partial_t+\bmath{u}\cdot\nabla
\end{equation}
is the Lagrangian derivative, $\bmath{u}$ is the velocity, $p$ is the pressure and $\rho$ is the density. The tidal potential is expanded as
\begin{equation}
    \label{ring_model_eq:tidal_potential}
    \Phi_t = -\Omega_0 S_0 x^2+\frac{1}{2}\nu_0^2 z^2,
\end{equation}
where $S_0 = -(r d\Omega/dr)_0$ is the orbital shear rate and $\nu_0^2 = (\partial_{zz} \Phi)_0$ is the square of the vertical oscillation frequency of a test particle perturbed from its circular orbit. Similarly, inertial restorative forces cause a natural radial oscillation about this orbit which is defined by the epicyclic frequency $\kappa_0$ given by,
\begin{equation}
    \label{ring_model_eq: epicyclic_frequency}
    \kappa_0^2 = 2\Omega_0(2\Omega_0-S_0).
\end{equation}
Henceforth we will drop the subscript on the orbital velocity, shear rate, and vertical/epicyclic frequencies in order to simplify our notation. We restrict our basic model to an isentropic energy equation with adiabatic index $\gamma$ but allow for compressibility. Density $\rho$ and pressure $p$ are then governed by
\begin{gather}
    \label{ring_model_eq:mass_continuity}
    D\rho = -\rho \Delta, \\
    \label{ring_model_eq:energy}
    Dp = -\gamma p \Delta,
\end{gather}
where
\begin{equation}
    \label{ring_model_eq:divergence}
    \Delta = \nabla\cdot \bmath{u} 
\end{equation}
is the velocity divergence. We look for axisymmetric dynamical solutions for which the density and pressure are described by a common materially invariant function $f(x,z,t)$. This allows us to perform the separation of variables
\begin{gather}
    \label{ring_model_eq:rho_separation}
    \rho = \hat{\rho}(t)\Tilde{\rho}(f), \\
    \label{ring_model_eq:p_separation}
    p = \hat{p}(t)\Tilde{p}(f).
\end{gather}
We then enforce linear flow fields, which capture the lowest order global motions supported by the tori, such that
\begin{equation}
    \label{eulerian_eq:linear_flow}
    u_i = A_{ij}x_j,
\end{equation}
where $A_{ij}$ is a time-dependent, square \textit{flow matrix}. Since this linear flow maps ellipses to ellipses, the materially conserved function $f$ should be a quadratic function of the coordinates such that
\begin{equation}
    \label{ring_model_eq:elliptical_f}
    f = C-\frac{1}{2}S_{ij}x_i x_j,
\end{equation}
where $C$ is some constant and $S_{ij}(t)$ is a time dependent, positive-definite \textit{shape matrix} with $S_{i2}=S_{2i}=0$ in the $y$-independent case. Thus contours of equal density and pressure trace out elliptical contours, which are described by the time evolution of the shape matrix. Imposing the material conservation condition for $f$ at all points in space requires that
\begin{equation}
    \label{ring_model_eq: shape_matrix_evolution}
    d_t S_{ij} + S_{ik}A_{kj}+S_{jk}A_{ki}=0,
\end{equation}
which gives three independent ODEs for $S_{11}$, $S_{13}$ and $S_{33}$. Meanwhile, the flow matrix evolution is deduced by inserting our assumptions into the equation of motion and gathering terms linear in each spatial coordinate. We must also impose $d\Tilde{p}/df = \Tilde{\rho}$ so that the pressure gradient term is compatible with this linear form in the coordinates. This gives rise to six ODEs,
\begin{gather}
    \label{eulerian_eq:A11}
    d_t A_{11}+A_{11}^2+A_{13}A_{31}-2\Omega A_{21} =2\Omega S+\hat{T}S_{11}, \\
    \label{eulerian_eq:A13}
    d_t A_{13}+A_{11}A_{13}+A_{13}A_{33}-2\Omega A_{23}=\hat{T}S_{13}, \\
    \label{eulerian_eq:A21}
    d_t A_{21}+A_{21}A_{11}+A_{23}A_{31}+2\Omega A_{11} =0, \\
    \label{eulerian_eq:A23}
    d_t A_{23}+A_{21}A_{13}+A_{23}A_{33}+2\Omega A_{13} = 0, \\
    \label{eulerian_eq:A31}
    d_t A_{31}+A_{31}A_{11}+A_{33}A_{31}=\hat{T}S_{13}, \\
    \label{eulerian_eq:A33}
    d_t A_{33}+A_{31}A_{13}+A_{33}^2=-\nu^2+\hat{T}S_{33},
\end{gather}
where $\hat{T}(t) = \hat{p}/\hat{\rho}$ is a characteristic temperature. This evolves according to 
\begin{equation}
\label{eulerian_eq:t_hat_evolution}
    d_t{\hat{T}} = -(\gamma-1)\hat{T}\Delta,
\end{equation}
where $\Delta = A_{11}+A_{33}$ is the velocity divergence.
\subsection{Lagrangian perspective}
This model may alternatively be reformulated from a Lagrangian perspective. We construct a material mapping of points from an arbitrary, stationary reference state, denoted by $\bmath{x}_0 = (x_0,y_0,z_0)$, to the dynamical state $\bmath{x}$ by means of the linear transformation
\begin{equation}
    \label{ring_model_eq:jacobian_transformation}
    \bmath{x}_0 \mapsto \bmath{x}: \quad x_i = J_{ij} x_{0,j},
\end{equation}
where $J_{ij}$ is the time dependent Jacobian matrix. For the assumed axisymmetric setup $J_{12} = J_{32} = 0$ and $J_{22}=1$, so the 6 remaining independent components describe the linear flow field $u_i = \dot{J}_{ij}x_j$. We load mass in the reference state such that the materially conserved density and pressure contours lie on circles with radius $L R$. Here, $R = \sqrt{2(C-f)}$ is a dimensionless radius measured in units of the characteristic length $L$, which arises when taking the second mass weighted moment of the reference distribution. Our separation of variables then becomes
\begin{equation}
    \label{ring_model_eq: rho_p_reference_state}
    \rho_0(\bmath{x}_0) = \hat{\rho}_0\Tilde{\rho}(R(\bmath{x}_0)),\quad \text{and} \quad
    p_0(\bmath{x}_0) = \hat{p}_0\Tilde{p}(R(\bmath{x}_0)),
\end{equation}
where $\rho_0$ and $p_0$ denote the density and pressure in the reference state whilst $\hat{\rho}_0$ and $\hat{p}_0$ are characteristic density and pressure factors. 

In Paper I we outline the construction of a Lagrangian composed of the kinetic, rotational, internal and potential energies
\begin{align}
    \label{lagrangian_eq:total_lagrangian}
    \mathcal{L} &= \frac{1}{2}(\dot{J}_{11}^2+\dot{J}_{13}^2+\dot{J}_{21}^2+\dot{J}_{23}^2+\dot{J}_{31}^2+\dot{J}_{33}^2)-\frac{\hat{T}_0}{(\gamma-1) J^{\gamma-1} L^2} \nonumber \\
    &+\Omega S (J_{11}^2+J_{13}^2)-\frac{1}{2}\nu^2(J_{31}^2+J_{33}^2)+2\Omega(J_{11}\dot{J}_{21}+J_{13}\dot{J}_{23}),
\end{align}
where $J = \text{det}(J_{ij}) = J_{11}J_{33}-J_{13}J_{31}$ is proportional to the area of the elliptical cross-section of the ring and $\hat{T}_0=\hat{p}_0/\hat{\rho}_0$ is a characteristic temperature. The usual Euler-Lagrange equations then give the dynamical equations
\begin{align}
    \label{lagrangian_eq:J11_ode}
    & \ddot{J}_{11} = 2\Omega\dot{J}_{21}+2\Omega S J_{11}+\frac{\hat{T}_0}{J^\gamma L^2}J_{33}, \\
    \label{lagrangian_eq:J13_ode}
    & \ddot{J}_{13} = 2\Omega\dot{J}_{23}+2\Omega S J_{13}-\frac{\hat{T}_0}{J^\gamma L^2}J_{31}, \\
    \label{lagrangian_eq:J21_ode}
    & \ddot{J}_{21} = -2\Omega\dot{J}_{11}, \\
    \label{lagrangian_eq:J23_ode}
    & \ddot{J}_{23} = -2\Omega\dot{J}_{13}, \\
    \label{lagrangian_eq:J31_ode}
    & \ddot{J}_{31} = -\nu^{2}J_{31}-\frac{\hat{T}_0}{J^\gamma L^2}J_{13}, \\
    \label{lagrangian_eq:J33_ode}
    & \ddot{J}_{33} = -\nu^{2}J_{33}+\frac{\hat{T}_0}{J^\gamma L^2}J_{11}.
\end{align}
The conservation of angular momentum gives rise to the integrability of equations \eqref{lagrangian_eq:J21_ode} and \eqref{lagrangian_eq:J23_ode} which allows us to reduce this system to 4 second-order, coupled ODEs,
\begin{align}
   & \ddot{J}_{11} +\kappa^2 J_{11}=2C_z+\frac{\hat{T}_0}{J^\gamma L^2}J_{33}, \label{nonlinear_eq: J_11_ode}\\
   & \ddot{J}_{13} +\kappa^2 J_{13}=2C_x-\frac{\hat{T}_0}{J^\gamma L^2}J_{31}, \label{nonlinear_eq: J_13_ode}\\
   & \ddot{J}_{31} +\nu^2 J_{31}=-\frac{\hat{T}_0}{J^\gamma L^2}J_{13}, \label{nonlinear_eq: J_31_ode}\\
   & \ddot{J}_{33} +\nu^2 J_{33}=\frac{\hat{T}_0}{J^\gamma L^2}J_{11}, \label{nonlinear_eq: J_33_ode}
\end{align}
where $C_x$ and $C_z$ represent the constants arising from circulation conservation and we have made use of the definition of the epicyclic frequency $\kappa$ to eliminate the shear rate $S$. 

\subsection{Physical interpretation and connection with warping}
\label{section:ring_model:interpretation}
It is worth emphasising the physical intuition behind these variables. $J_{11}$ and $J_{33}$ describe a radial and vertical stretching of the ring respectively, capable of capturing breathing motions. $J_{13}$ corresponds to the vertical shear of horizontal flows whilst $J_{31}$ describes the ring tilting as one moves radially outwards (for helpful visualisations refer to Paper I). The Lagrangian form of the equations clearly elucidates the oscillatory structure underlying the ring system. The left-hand side terms correspond to free harmonic oscillators, whilst on the right-hand side, matters are complicated by the pressure terms which couple the oscillators together.

As discussed in Paper I, the off-diagonal Jacobian elements act to break the midplane symmetry of the elliptical rings and can be identified with bending modes. Indeed, these tilting motions, as observed within the shearing box orbital frame, can be reinterpreted in a global, non-rotating reference frame as a series of nested circular orbits with a radially dependent inclination. This tilting of streamlines may be thought of as an $m=1$ azimuthal mode and hence associated with a warped structure \citep[]{Ogilvie2013}. To illustrate this, imagine setting up a line of test particles on circular orbits with a radial, linear variation in inclination about the reference shearing box orbital plane. In a Keplerian potential these orbits are closed and describe a fixed, warped annulus. However, when viewed from the rotating shearing box frame, the line of test particles rock up and down, simply tracking the geometry of the tilted annulus. 

More generally, the inclusion of pressure in a gaseous disc couples these particle orbits and may introduce some precession of the streamlines. As the global structure slowly rotates, the oscillation period in the shearing box frame will depart from the orbital period. These two perspectives are connected by Doppler shifting the $m=1$ warping mode such that 
\begin{equation}
    \omega_p = \Omega-\omega,
\end{equation}
where $\omega_p$ is the precessional frequency in the global frame and $\omega$ is the frequency of the tilting mode in the local model. Thus $\omega_p<0$ and $\omega_p>0$ correspond to retrograde and prograde precessing warped structures respectively.  

In order to quantitatively connect this with global warped theory, we will introduce a local measure of the warp amplitude. Consider a test particle on an inclined orbit such that it undergoes vertical oscillations in the local model according to $z = \Re(Ze^{-i\Omega t})$ where $Z$ is a complex amplitude. Then the magnitude of $Z$ is proportional to the orbit inclination whilst the argument is related to the longitude of ascending node. Thus we may express this quantity in terms of the classic complex tilt variable $W = l_x+i l_y$ such that $Z=-r_0 W$. Here $\bmath{l}$ is the unit tilt vector, pointing normal to the circular orbits of the test particle, which clearly encapsulates the amplitude and phase of the $z$ motion about the reference plane. The value of $Z$ is extracted from the local model via 
\begin{equation}
    Z = \left( z+\frac{i\dot{z}}{\Omega}\right)e^{i\Omega t}.
\end{equation}
Consider a set of particles along the midplane of the ring $z_0=0$ labelled by reference coordinate $x_0$, such that $z = J_{31}x_0$ and 
\begin{equation}
    Z = \left(J_{31}+\frac{i}{\Omega}\dot{J}_{31} \right)x_0 e ^{i\Omega t}.
\end{equation}
Then along this line $x=J_{11}x_0$ such that the warp amplitude, defined as the gradient $\psi \equiv dZ/dx$, is given by
\begin{equation}
    \label{eq:ring_model:psi}
    \psi = \frac{1}{J_{11}}\left(J_{31}+\frac{i}{\Omega}\dot{J}_{31}\right)e^{i\Omega t}.
\end{equation}
This connection between the local tilting modes and the global warped perspective is crucial for understanding the solutions derived later.
\section{Numerical motivation}
\label{section:numerical_motivation}

The simplicity of the linear harmonic form presented by the left hand side of equations \eqref{nonlinear_eq: J_11_ode} -- \eqref{nonlinear_eq: J_33_ode} makes them an attractive framework to explore the nonlinear effects introduced by the right hand side pressure terms. The large number of degrees of freedom and significant nonlinearity introduced by the pressure couplings means it is instructive to first numerically solve this system of ODEs. This will reveal a rich range of dynamical behaviour. Using an implicit Runge-Kutta integrator we test a range of tilted initial conditions which break the midplane symmetry of the ring. As we increase the amplitude of the tilt and depart further from equilibrium, we identify two distinct nonlinear warping regimes which will motivate our analysis in subsequent sections.

\subsection{Tilting setup}
\label{subsection:tilting_setup}
As demanded by the gap in the current warped disc theory, we will focus on the resonant regime for which the epicyclic, vertical and orbital frequencies are all equal with $\kappa=\nu=\Omega$. Without loss of generality we can choose our units such that $\Omega =1$ and $L=1$. We assume a typical adiabatic index $\gamma = 5/3$ and set up a thin equilibrium ring with $J_{11} = 100$ and $J_{33}=1$ such that the aspect ratio is given by $\epsilon = J_{33}/J_{11} = 0.01$. This choice ensures that the length scale of the warp is much longer that the disc scale-height. As described in Paper I, the vertical equilibrium is established via the hydrostatic balance described by equation \eqref{nonlinear_eq: J_33_ode} which sets the value of $\hat{T}_0 = \epsilon J^{\gamma}$. The finite width of the ring then incurs a radial pressure gradient which is balanced by an enhanced shear. This manifests as a reduced value of the Bjerknes circulation constant $C_{z} = (1/2)J_{11}(1-\epsilon^2)$, as the local shear flow vorticity component counteracts the global rotational vorticity.

In Paper I, we investigated linear tilting modes by slightly perturbing this equilibrium ring and found close correspondence with linear bending-wave theory. We now gain a foothold on the transition to nonlinear tilting dynamics by releasing the ring from increasing tilt angles $\theta_{t}$. We simply rotate the equilibrium ring so that $\theta_{t}$ corresponds to the angle between the ellipse's major axis and reference plane measured in radians. Releasing from this rotated state presents a general configuration which naturally engages the warping motions of interest. In order to interpret the change in the dynamics as the amplitude is increased we will examine the warp amplitude $\psi$ as defined in equation \eqref{eq:ring_model:psi}. This is a useful diagnostic for understanding the tilting and precession of the ring as the warped structure evolves. Linear bending waves generally trace out elliptical paths in a polar plot of $\psi$. As the amplitude of the tilting perturbation increases we expect the nonlinearities to significantly distort this picture, as we shall soon see.
\begin{figure}
    \centering
    \includegraphics[width=0.92\columnwidth]{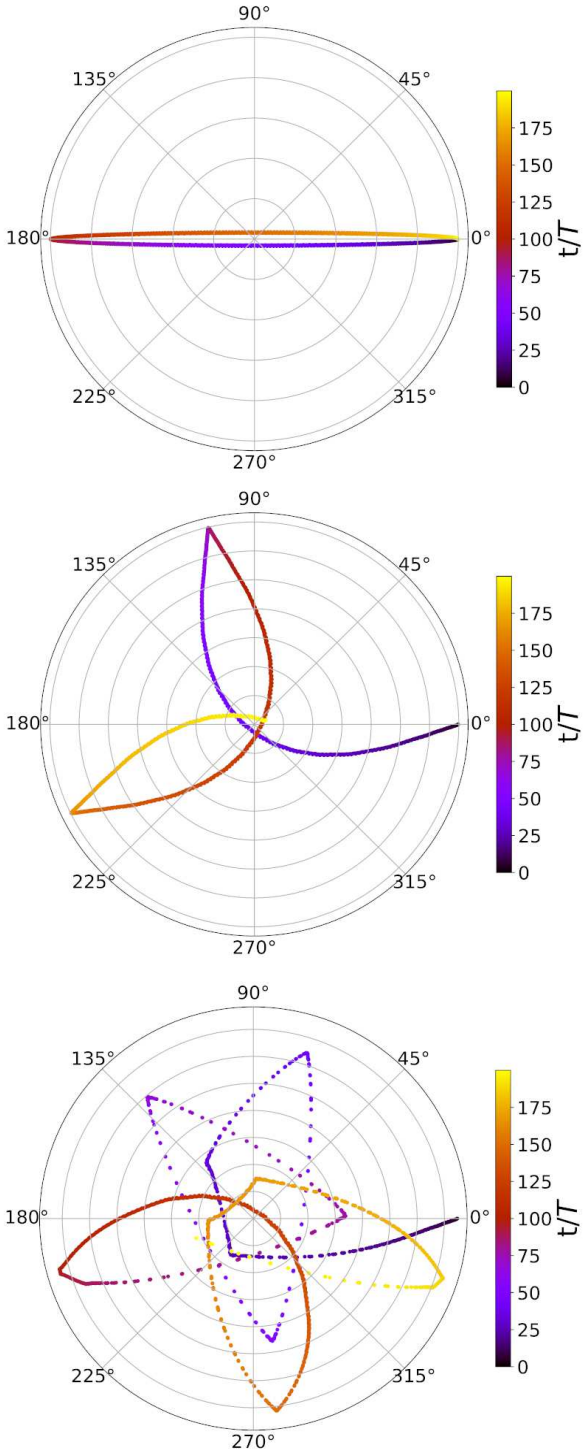}
    \caption{The polar plots of the warp amplitude $\psi$ tracked over $200$ orbital periods for different initial tilt angles. The complex nature of this variable means that the plots display the evolution of both the magnitude and the phase of the warp. These govern the linear rate at which fluid streamlines are tilted from the reference plane when moving radially, and the global orientation of this tilting, respectively. \textit{Upper panel}: $\theta_t = 0.01$ is a small tilt. The elliptical track is indicative of the linear bending mode regime. \textit{Middle panel}: $\theta_t = 0.14$ is a moderate tilt. The elliptical track is smoothly distorted as amplitude and phase of tilt and shear oscillators are modulated on secular timescales. \textit{Lower panel}: $\theta_t = 0.15$ is a critical tilt. The smooth track suddenly changes behaviour as the vertical oscillator is resonantly driven into a bouncing regime and feedback onto the warp occurs impulsively.}
    \label{fig:warp_amplitude}
\end{figure}
%
\subsection{Transition to nonlinearity}

For small amplitude $\theta_{t}$, the $J_{13}$ and $J_{31}$ oscillators exhibit a beating pattern as both the in-phase and anti-phased linear tilting modes are excited by a general initial condition. This corresponds to elliptical tracks traced out by the warp amplitude as seen in Paper I. Here we observe that for a tilt angle of $\theta=0.01$, $\psi$ traces a squashed elliptical track as seen in the upper panel of Fig.~\ref{fig:warp_amplitude}. This path is indicative of the secular precession of the tilted ring structure over many orbital timescales. As the initial tilt amplitude is increased, the system smoothly extends into the nonlinear regime. Whilst the shear and tilt oscillators remain largely harmonic in their behaviour, the vertical oscillator $J_{33}$ is driven to nonlinear amplitudes and becomes dynamically important. This nonlinearity feeds back onto the warp, driving a slow modulation of the phase and amplitude of the tilt and shear oscillators. This distorts the linear warp amplitude elliptical tracks into more interesting configurations, as shown for the $\theta_t = 0.14$ run in the middle panel of Fig.~\ref{fig:warp_amplitude}. The Jacobian variables for this run are plotted in the upper four panels of Fig.~\ref{fig:ring_solver_run}.

\subsection{Critical onset of bouncing regime}

\begin{figure*}
\centering
    \includegraphics[width=\textwidth]{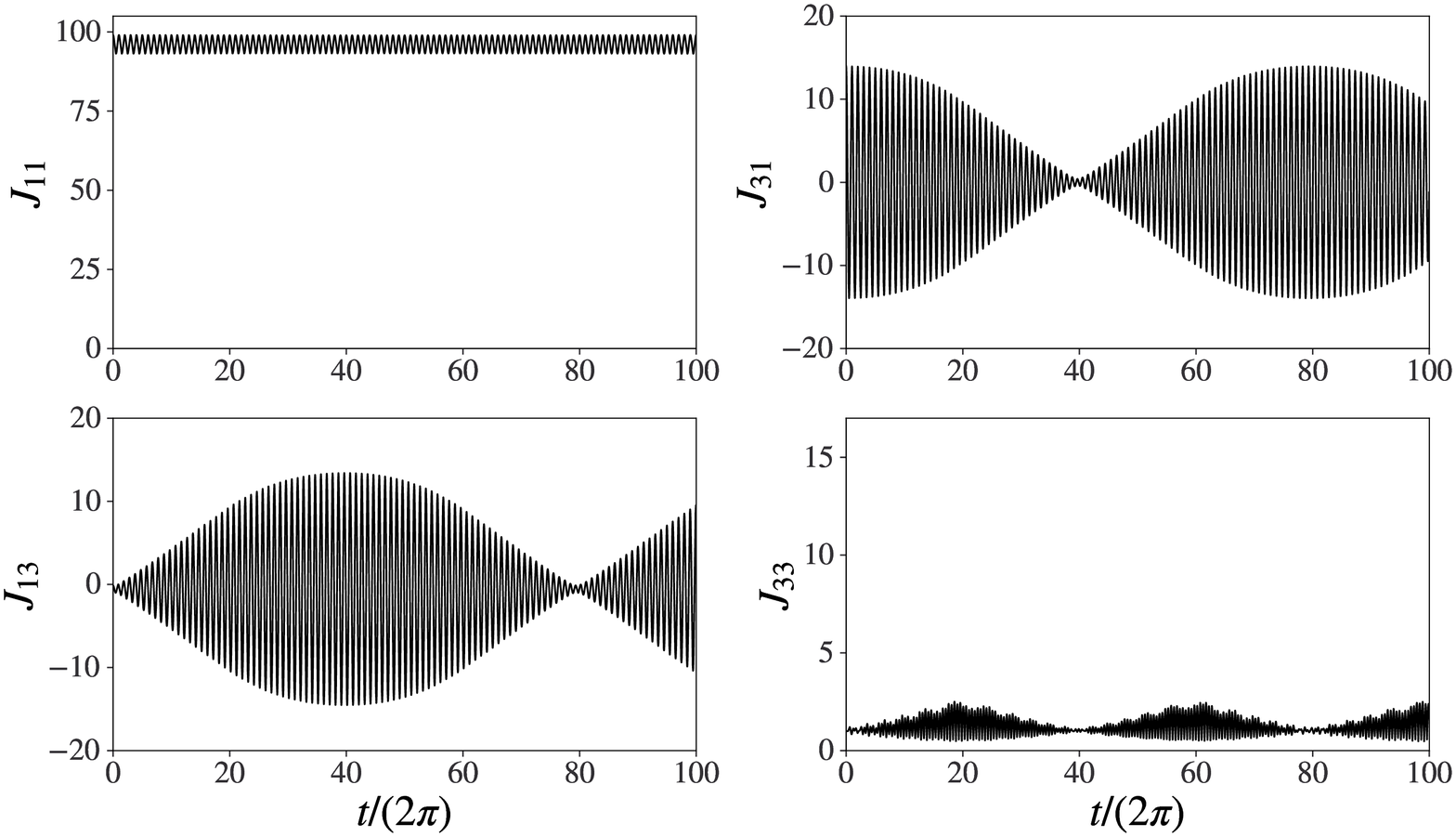}
    \includegraphics[width=\textwidth]{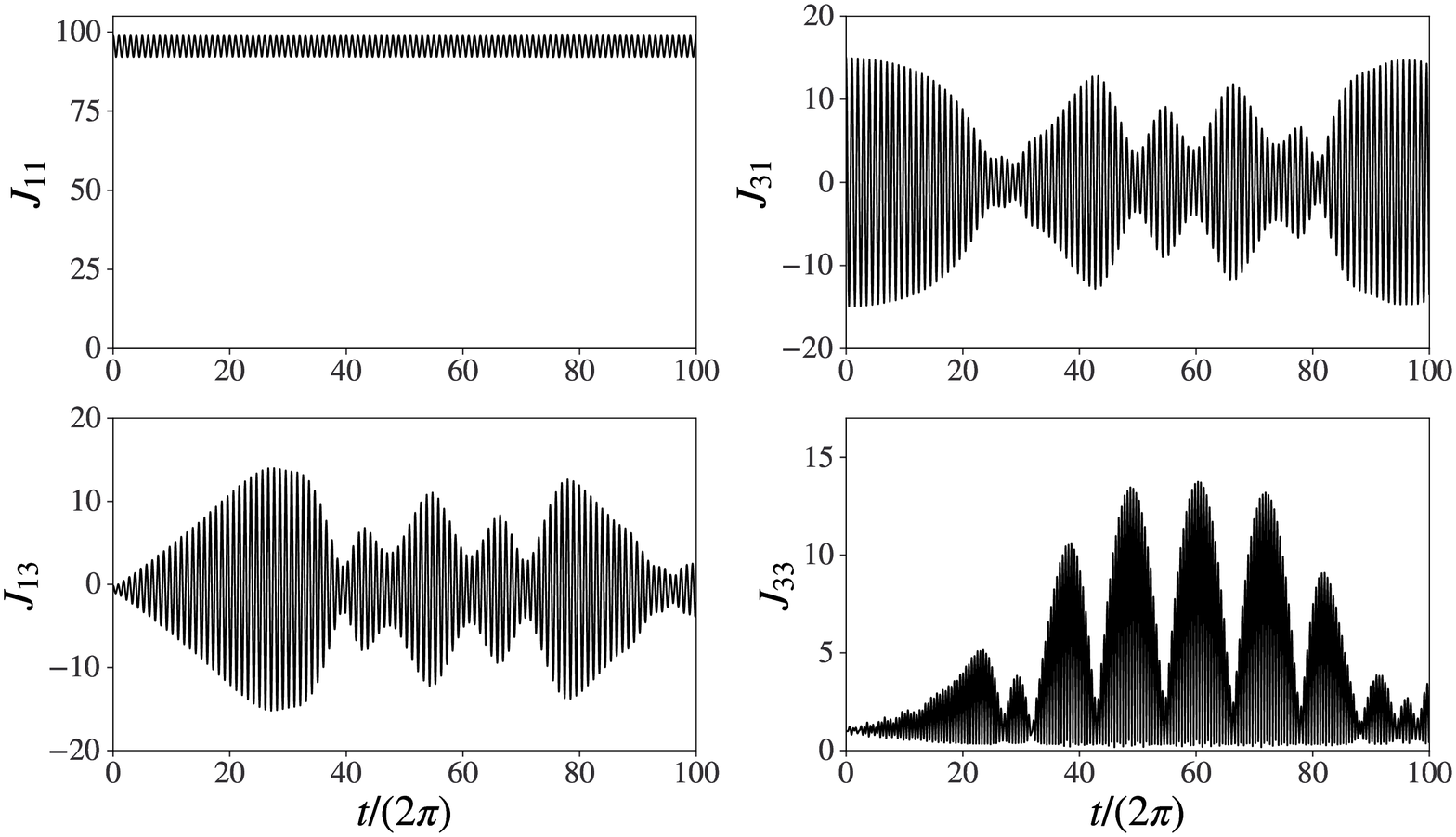}
\caption{The numerically integrated solution to equations \eqref{nonlinear_eq: J_11_ode} -- \eqref{nonlinear_eq: J_33_ode} for resonant runs with $\kappa=\nu=\Omega=1$ and $\gamma = 5/3$. The ring is initialised with aspect ratio $\epsilon=0.01$ and then rotated from this equilibrium state by $\theta_{t}$. This measures the angle between the major axis of the ellipse and the x-axis. The $J_{13}$ and $J_{31}$ panels encapsulate the shear and tilting motions respectively whilst the $J_{33}$ panel captures the compressive breathing motions. \textit{Upper four panels:} $\theta_t = 0.14$ corresponding to the middle panel of Fig.~\ref{fig:warp_amplitude} shows the smooth modulation regime. \textit{Lower four panels:} $\theta_t = 0.15$ corresponding to the lower panel of Fig.~\ref{fig:warp_amplitude} shows extreme vertical bouncing motions.}
\label{fig:ring_solver_run}
\end{figure*}

When $\theta_{t}\gtrapprox 0.15$ a dynamically distinct behaviour arises. Note that this critical angle generally depends on the system parameters i.e. $\epsilon$ and $\gamma$. The lower panel in Fig.~\ref{fig:warp_amplitude} plots $\psi$ when the ring is released from this initial tilt and shows a rapid, possibly chaotic evolution. To gain further insight, the individual Jacobian components are plotted in the lower four panels of Fig.~\ref{fig:ring_solver_run}. Comparison with the $\theta_{t}=0.14$ run in the upper four panels emphasises a drastically different behaviour. This demonstrates a resonant coupling between the tilting motions associated with the $J_{31}$ and $J_{13}$ components and the breathing motions indicated by the $J_{33}$ component. We see that the initial beating envelopes of the tilt and shear terms are disrupted as their combined effect drives a growth in the breathing motion. This is shown by the large amplitude, compressive bumps in the lower right panel. The system enters into a quasi-periodic regime with strong mode coupling between the vertical breathing and warping oscillations. The driving of such extreme breathing modes has also been separately recognised in the periodically forced scale heights associated with elliptical fluid flows \citep{Ogilvie2014}.

\begin{figure}
     \centering
         \includegraphics[width=\columnwidth]{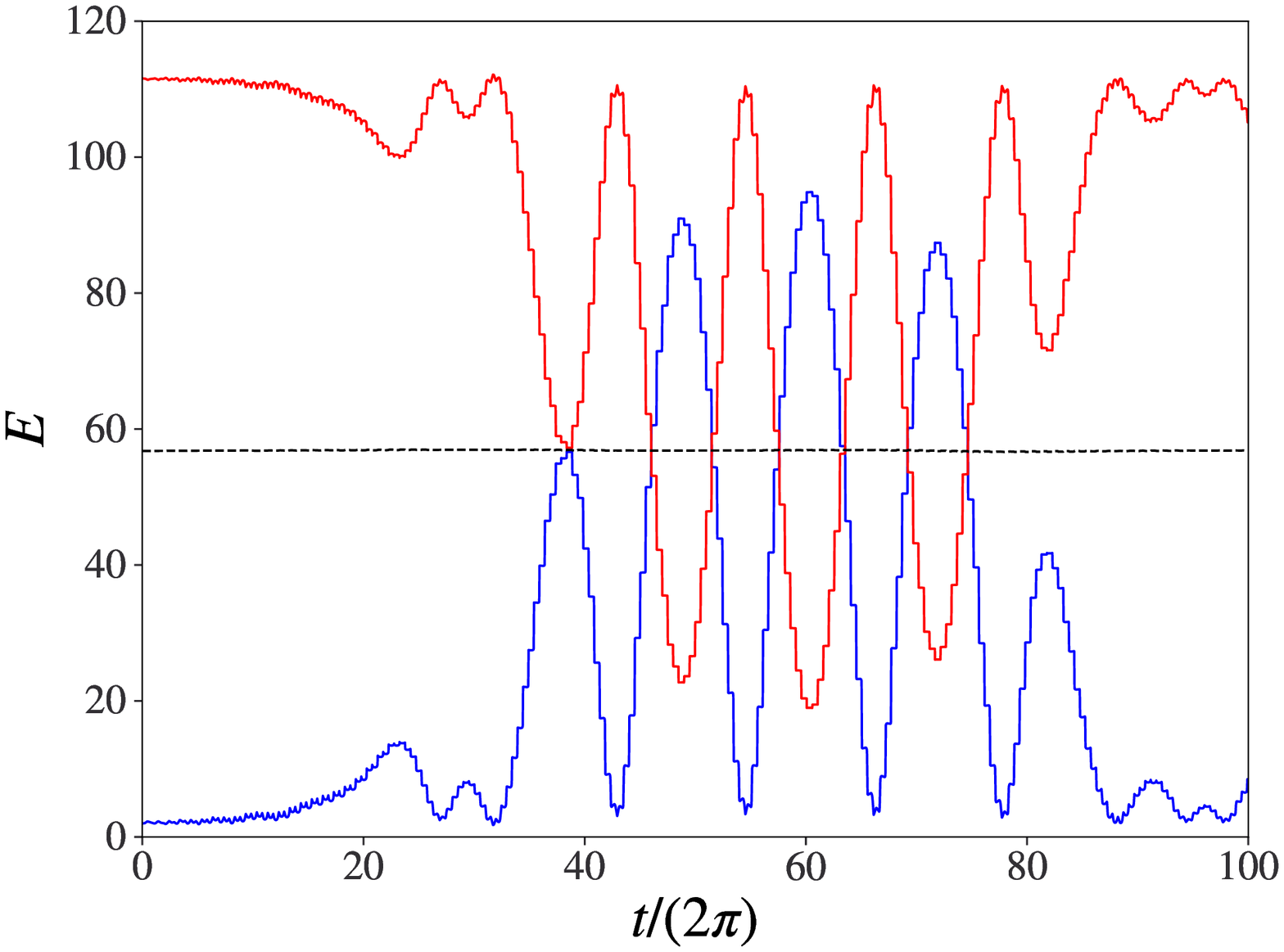}
         \caption{Comparison of the energy partitioning between the breathing and warping motions for the run with $\theta_t=0.15$. The red line plots the tilting energy contribution appearing in the Lagrangian, $E_{\text{tilt}} = \frac{1}{2}(\dot{J}_{13}^2+\dot{J}_{31}^2+J_{13}^2+J_{31}^2)-2C_x J_{13}$. The blue line plots the vertical breathing contributions $E_{\text{breathe}} = \frac{1}{2}(\dot{J}_{33}^2+J_{33}^2)+\frac{\hat{T}_0}{(\gamma-1)J^{\gamma-1}}$. The dashed black line plots the average between these two energies $(\mathcal{L}_{\mathsf{tilt}}+\mathcal{L}_{\mathsf{breath}})/2.$}
         \label{fig:energy_partition}
\end{figure}

Whilst in the smooth nonlinear regime the breathing and warping modes remain largely disconnected reservoirs of energy, in this compressive nonlinear phase the pressure couplings facilitate a large energy exchange flowing back and forth between these motions. This is visualised clearly in Fig.~\ref{fig:energy_partition} which shows how the energy is partitioned between the different modes over time. The red line plots the energy terms in the Lagrangian corresponding to the warping motions i.e. the kinetic and potential energies involving $J_{13}$ and $J_{31}$. Meanwhile, the blue line plots the kinetic and potential energies of the $J_{33}$ oscillator plus the contribution from the internal energy. It is natural to combine the internal energy with the breathing mode since only compressive motions can heat the ring. Indeed, in linear theory the tilting modes are incompressible and internal energy is conserved. We see that both lines are essentially symmetric about the average energy, denoted by the black dashed line. A large dip in warping energy is balanced by an increase in breathing energy and vice versa. This emphasises the mode coupling channel which is clearly active. Furthermore, we note the red and blue lines appear to vary in a step like manner. This is not an artefact of numerical resolution but in fact a key part of the dynamical behaviour. Each step coincides with a compression of the ring where $J_{33}$ is squashed. At these discrete times, the cross sectional area of the ring is small and the determinant value $J$ is minimised. It is at these instances that the pressure coupling terms on the right hand side of equations \eqref{nonlinear_eq: J_11_ode} -- \eqref{nonlinear_eq: J_33_ode} dominate and allow for an impulsive exchange of energy. It is this impulsive coupling mechanism that will motivate our analytical progress in this regime in the following sections.
%

\section{Smooth modulation theory}
\label{section:smooth_modulation}
\subsection{Asymptotic scalings}
We first confront the smooth nonlinear regime where we expect a secular modulation of the linear oscillatory solutions. For a thin ring we anticipate that the radial breathing motions are not dynamically important and so we ignore equation \eqref{lagrangian_eq:J11_ode} and set $J_{11}$ to be constant. For the equilibrium ring with small aspect ratio $\epsilon$ we have the characteristic temperature $\hat{T}_0 = \epsilon J^{\gamma}$. The scale invariance of ideal hydrodynamics means we are free to adopt a reference state with area of order unity, so we take the determinant $J \sim O(1)$ and $\hat{T}_0 = \epsilon$ which directly introduces a small parameter into the governing equations. To facilitate this area scaling we will take the radial and vertical deformations to be $J_{11} \sim 0(\epsilon^{-1/2})$ and $J_{33} \sim O(\epsilon^{1/2})$ respectively. We are interested in exploring nonlinear warps, so adopt the scalings $J_{13}\sim J_{31}\sim O(1)$ such that they contribute at leading order to $J$. Inserting these scalings into the reduced Lagrangian for the tilt, shear and vertical oscillators is
\begin{equation}
    L = \underbrace{\frac{1}{2}\left( \dot{J}_{13}^2+\dot{J}_{31}^2-J_{13}^2-J_{31}^2 \right)}_{L_0 \sim O(1)} +\underbrace{\frac{1}{2}\left( \dot{J}_{33}^2 -J_{33}^2-\frac{\epsilon}{(\gamma-1)J^{\gamma-1}}\right)}_{\epsilon L_1 \sim O(\epsilon)}.
\end{equation}
We see that the Lagrangian is split into a leading order component which just describes harmonic motion of the tilt and shear. At higher order we see the contribution from the vertical oscillator kinetic, potential and internal energies. Notably the tilt and shear are coupled to the vertical oscillator through the internal energy term and will drive a slow modulation of the harmonic motion phase and amplitude over longer timescales. To capture the fast harmonic motion and the slow evolution owing to the nonlinearities we introduce the multiple timescales expansion
\begin{align}
J_{11} &= \epsilon^{-1/2}, \label{eq:modulation:J_11_scaling}\\
J_{13} &= J_{13,0}(t,T)+\epsilon J_{13,1}(t,T)+O(\epsilon^2), \\
J_{31} &= J_{31,0}(t,T)+\epsilon J_{31,1}(t,T)+O(\epsilon^2), \\
J_{33} &= \epsilon^{1/2}J_{33,0}(t,T)+O(\epsilon^{3/2}),
\end{align}
where $T = \epsilon t$ is a slow timescale treated as an independent parameter. Thus the full time derivatives become
\begin{equation}
    d_t = \partial_t + \epsilon\partial_T, \quad d_{tt} = \partial_{tt} +2\epsilon\partial_{tT}+\epsilon^2\partial_{TT}.
\end{equation}
Inserting this expansion into the dynamical equations \eqref{nonlinear_eq: J_13_ode}--\eqref{nonlinear_eq: J_33_ode} yields a hierarchy of equations ordered in powers of $\epsilon$. One should note that the scale invariance of the equations of motion means that we are in fact free to stretch the results provided the underlying aspect ratio is preserved. This scale invariance may be parameterised relative to the width of the ring $J_{11}$ so the dynamics is similar if we re-scale variables such that $J_{13}$ and $J_{31}$ are of order $O(\epsilon^{1/2} J_{11})$, whilst $J_{33}\sim O(\epsilon J_{11})$. This is important to remember later on when comparing our theory to general numerical runs where the scaling of the elliptical area measure $J$ is not necessarily of order unity. 

\subsection{Modulation equations}

As anticipated, at leading order $O(\epsilon)$ we have 
\begin{equation}
    \partial_{tt} J_{13,0}+J_{13,0} = 0 \quad \text{and} \quad \partial_{tt} J_{31,0}+J_{31,0} = 0.
\end{equation}
These have harmonic solutions 
\begin{equation}
    J_{13,0} = \Re\left[ A(T)e^{-it}\right] \quad \text{and} \quad J_{31,0} = \Re\left[B(T)e^{-it}\right],
\end{equation}
where $\Re$ denotes the extraction of the real part and $A(T)$ and $B(T)$ are complex amplitudes encoding the slow modulation of oscillator amplitude and phase. At order $O(\epsilon^{1/2})$ we obtain the leading order equation for the vertical oscillator
\begin{equation}
    \label{equation:weaklynonlinear:J330}
    \partial_{tt} J_{33,0}+J_{33,0} = \frac{1}{H^\gamma},
\end{equation}
where $H \equiv J_{33,0}-J_{13,0}J_{31,0}$. This equation may be tackled by changing variables in favour of $H$ such that
\begin{equation}
    \label{eq:weaklynonlinear:vertical_oscillator}
    \partial_{tt} H+H-\frac{1}{H^\gamma} = -\partial_{tt}(J_{13,0}J_{31,0})-J_{13,0}J_{31,0},
\end{equation}
showing that the compressional motion is driven by the product of the tilt and shear. The right-hand side of this equation is periodic with frequency 2 (i.e. twice the orbital frequency). Periodic solutions with frequency 2 are possible for a certain range of forcing amplitudes, as we shall discuss in Section \ref{subsection:forced_vertical} below. We assume here that the solutions are indeed periodic in t, rather than the more general quasi-periodic solutions that include a free oscillation as well as the forced one. In the meantime we will expand to next order in the aspect ratio hierarchy so at $O(\epsilon^1)$ we have
\begin{align}
    \label{eq:weaklynonlinear:J131}
    \partial_{tt} J_{13,1} +J_{13,1} &= -2 \partial_{tT} J_{13,0} - \frac{J_{31,0}}{H^\gamma} \equiv f_{13,1}, \\
    \label{eq:weaklynonlinear:J311}
    \partial_{tt} J_{31,1} +J_{31,1} &= -2 \partial_{tT} J_{31,0} - \frac{J_{13,0}}{H^\gamma} \equiv f_{31,1}.
\end{align}
On the left hand side we see the linear operator $\partial_{tt}+1$ which yields complementary harmonic solutions. We will also denote the right hand side forcing terms as $f_{13,1}$ and $f_{31,1}$. A necessary condition for periodic solutions requires that the forcing on the right hand side contains no $e^{\pm it}$ resonant Fourier components. This is equivalent to the Fredholm solvability conditions
\begin{equation}
    \left\langle f_{13,1} e^{\pm i t}\right\rangle = \left\langle f_{31,1} e^{\pm i t}\right\rangle = 0,
\end{equation}
where
\begin{equation}
    \left\langle \cdot \right\rangle \equiv \frac{1}{2\pi} \int_0^{2\pi} \cdot\, dt
\end{equation}
denotes averaging over the fast orbital timescale. Evaluating these conditions yields
\begin{equation}
    \label{eq:weaklynonlinear:solvability}
    -i \partial_T A = -\left\langle \frac{J_{31,0}}{H^\gamma}e^{it}\right\rangle 
    \quad \text{and} \quad 
    -i \partial_T B = -\left\langle \frac{J_{13,0}}{H^\gamma}e^{it}\right\rangle
\end{equation}
which gives the evolution of the complex amplitudes over secular timescales based on the fast averaging of the lower order equations.

\subsection{Averaged Lagrangian formulation}

Since we are dealing with ideal hydrodynamics as derived from a variational principle, we anticipate that the averaged terms can in fact be related to the averaged Lagrangian. This idea was first introduced by \cite{Whitham1965} with application to wave trains propagating through a slowly varying background and has applications in a wide variety of contexts. Returning to the forced vertical oscillator described by equation \eqref{equation:weaklynonlinear:J330} we see this can be derived from a Lagrangian 
\begin{equation}
    \label{eq:weaklynonlinear:L_10}
    L_{10} = \frac{1}{2}(\dot{J}_{33,0}^2-J_{33,0}^2) -\frac{H^{-(\gamma-1)}}{\gamma-1}, 
\end{equation}
which is the leading order $O(\epsilon^1)$ contribution from the vertical part of the full Lagrangian. The Lagrangian explicitly depends on time and the forcing parameters $A$ and $B$ through the product of tilt and shear oscillations appearing in $H$, and implicitly through the dependence of the $J_{33,0}$ solution as forced by the warp. These complex amplitudes encode two degrees of freedom each, encapsulating the amplitude and phase, so the complex conjugated quantities $\Bar{A}$ and $\Bar{B}$ may also be treated as independent quantities. Thus consider $L_{10} = L_{10}\left(A, \Bar{A}, B, \Bar{B}\right)$ and first compute
\begin{equation}
    \frac{\partial L_{10}}{\partial \Bar{A}} = -\frac{H^{-\gamma}}{2}e^{it}J_{31,0}+H^{-\gamma}\frac{\partial J_{33,0}}{\partial \bar{A}}+\dot{J}_{33,0}\frac{\partial \dot{J}_{33}}{\partial \bar{A}}-J_{33,0}\frac{\partial J_{33,0}}{\partial \bar{A}}.
\end{equation}
Using equation \eqref{equation:weaklynonlinear:J330} to replace $H^{-\gamma}$ in the second right-hand side term and averaging over the orbital period yields
\begin{equation}
    \frac{\partial}{\partial \bar{A}}\left\langle L_{10} \right\rangle = -\left\langle \frac{H^{-\gamma}}{2}e^{it}J_{31,0} \right\rangle + \left\langle \dot{J}_{33,0}\frac{\partial \dot{J}_{33,0}}{\partial \bar{A}}+\ddot{J}_{33,0}\frac{\partial J_{33,0}}{\partial \bar{A}}\right\rangle.
\end{equation}
Integrating by parts shows that 
\begin{equation}
    \left\langle \dot{J}_{33,0}\frac{\partial \dot{J}_{33,0}}{\partial \bar{A}}+\ddot{J}_{33,0}\frac{\partial J_{33,0}}{\partial \bar{A}}\right\rangle = 0,
\end{equation}
and so we have 
\begin{equation}
    \frac{\partial}{\partial \bar{A}}\left\langle L_{10} \right\rangle = -\left\langle \frac{H^{-\gamma}}{2}e^{it}J_{31,0} \right\rangle.
\end{equation}
Similarly we find that
\begin{equation}
    \frac{\partial}{\partial \bar{B}}\left\langle L_{10} \right\rangle = -\left\langle \frac{H^{-\gamma}}{2}e^{it}J_{13,0} \right\rangle.
\end{equation}
These can be inserted into the solvability conditions given by equation \eqref{eq:weaklynonlinear:solvability}, which then read
\begin{align}
    \label{eq:weaklynonlinear:Amodulation}
    \partial_T A &= 2i\frac{\partial}{\partial \Bar{A}} \left\langle L_{10} \right\rangle \\
    \label{eq:weaklynonlinear:Bmodulation}
    \partial_T B &= 2i\frac{\partial}{\partial \Bar{B}} \left\langle L_{10} \right\rangle.
\end{align}
These may be identified as the Euler-Lagrange equations for the orbital period time-averaged Lagrangian at leading order
\begin{align}
    \left\langle L \right\rangle &= \frac{1}{2\pi}\int_0^{2\pi} L_0(A, B, \partial_T A, \partial_T B ; c.c.) +\epsilon L_{1}(A, B ; c.c.)\,dt \nonumber \\
    & = 0 +\epsilon\left[ \frac{i}{4}(\Bar{A}\partial_T A-A\partial_T \Bar{A}+\Bar{B}\partial_T B - B\partial_T\Bar{B})+\left\langle L_{10} \right\rangle\right]+O(\epsilon^2),
\end{align}
where $c.c.$ denotes the complex conjugated variables. The variational principle for minimising the action $\int{\left\langle L\right\rangle} dT$ with respect to the generalised coordinates $(A,B,\partial_T A, \partial_T B;  c.c.)$, recovers equations \eqref{eq:weaklynonlinear:Amodulation} and \eqref{eq:weaklynonlinear:Bmodulation}. Alternatively we may identify $\langle \mathcal{H}\rangle=-\left\langle L_{10}\right\rangle$ as the Hamiltonian governing the secular evolution of the warp. Indeed, the Legendre transform of the averaged Lagrangian can be written as
\begin{equation}
    \left\langle\mathcal{H}\right\rangle = \partial_{T} A \frac{\partial \langle L \rangle}{\partial (\partial_T A)} + \partial_T B \frac{\partial \langle L \rangle}{\partial(\partial_T B)} + c.c. - \langle L \rangle.
\end{equation}
Since the Lagrangian is linear in the `velocity' coordinates, all terms cancel apart from $\langle L_{10} \rangle$ which is reversed in sign. In this case, the modulation equations formally have the structure of the complex Hamilton's equations
\begin{equation}
    iz = \frac{\partial \langle H \rangle}{\partial \bar{z}},
\end{equation}
where the canonical variable is identified as $z = A/\sqrt{2}$ or $B/\sqrt{2}$ for equations \eqref{eq:weaklynonlinear:Amodulation} and \eqref{eq:weaklynonlinear:Bmodulation} respectively. 

\subsection{Forced vertical oscillator}
\label{subsection:forced_vertical}

In order to ground this formalism, it remains to determine the evolution of the forced vertical oscillator at leading order $J_{33,0}$ so we can compute the averaged Lagrangian $\left\langle L_{10} \right\rangle$. Recall, the dynamics of the forced vertical oscillator is given by equation \eqref{eq:weaklynonlinear:vertical_oscillator}, where the forcing term on the right-hand side is given as a product of the tilt and shear
\begin{equation}
    f_{ts} \equiv J_{13,0}J_{31,0} = \Re\left[ A e^{-it}\right]\Re\left[ B e^{-it}\right].
\end{equation}
Expanding this forcing yields
\begin{equation}
    f_{ts} = \frac{1}{2}\Re\left\{|A||B|e^{-i[2t-\arg(A)-\arg(B)]}+A\Bar{B}\right\},
\end{equation}
where $|\cdot|$ and $\arg(\cdot)$ denote the modulus and argument respectively. We are free to choose the time origin since equation \eqref{eq:weaklynonlinear:vertical_oscillator} has no explicit temporal dependence. Taking $t \mapsto t-\arg(B)$ gives the forcing form
\begin{equation}
    f_{ts} = \frac{1}{2}\Re\left( Z_1 e^{-2it}+Z_1\right),
\end{equation}
where $Z_1 = A\Bar{B}$. The net forcing on the right hand side then becomes
\begin{equation}
    F \equiv -\partial_{tt}f_{ts} - f_{ts} = \Re\left(\frac{3}{2}Z_1 e^{-2it}-\frac{1}{2}Z_1\right).
\end{equation}
The solution for $H$ is therefore only dependent on the value of $Z_1$ so the modulation equations become
\begin{equation}
    \label{eq:weaklynonlinear:modulation_eqns_in_z}
    \partial_T A = 2i \frac{\partial \left\langle L_{10} \right\rangle}{\partial \bar{Z_1}}B \quad \text{and} \quad \partial_T B = 2i \frac{\partial \left\langle L_{10} \right\rangle}{\partial Z_1}A,
\end{equation}
where $\langle L_{10} \rangle$ is now regarded as a function of $Z_1$ and its complex conjugate. In order to find a solution for small $Z_1$, we first perform a weakly nonlinear analysis and find a series expansion solution for $H$. Taking $Z_1 = \delta Z_{1,0}$, with $\delta \ll 1$ and $Z_{1,0}\sim O(1)$, we expand the vertical oscillator equation in terms of
\begin{equation}
    H = H_{0}(t)+\delta H_{1}(t)+\delta^2 H_{2}(t)+\cdots,
\end{equation}
which again generates a hierarchy of equations. At leading order we recover the unforced vertical oscillator 
\begin{equation}
    \partial_{tt}H_0+H_0-H_0^{-\gamma} = 0.
\end{equation}
In order to conform with the $2\pi$ periodic boundary conditions required by the solvability conditions discussed previously, we will set the free oscillation to zero and adopt the equilibrium value $H_0 = 1$. At the $n^{th}$ order expansion we observe the general form 
\begin{equation}
    \left[\partial_{tt}+\omega_n(\gamma)^2\right]H_n(t) = F_{n}(\gamma,Z_1,t, H_{i<n}),
\end{equation}
where the forcing term on the right-hand side, $F_n$, depends on the lower order solutions. Again, we ignore the complementary solution so as to avoid quasiperiodic solutions and simply extract the forced oscillation at each order. This weakly nonlinear solution can be evaluated to arbitrary order and used to calculate the Lagrangian given by equation \eqref{eq:weaklynonlinear:L_10}. In terms of the $H$ variable this may be written as
\begin{equation}
    L_{10} = \frac{1}{2}\left( \partial_t H +\partial_t f_{ts}\right)^2 -\frac{1}{2}\left(H +f_{ts} \right)^2-\frac{H}{\gamma-1}\left(\partial_{tt}H +H -F\right).
\end{equation}
Computing the average then gives
\begin{align}
    \label{eq:weakly_nonlinear_averaged_L}
    \left\langle L_{10} \right\rangle = &-\frac{1+\gamma}{2(\gamma-1)}-\frac{1}{4}(Z_1+\Bar{Z_1}) \nonumber\\
    &-\gamma\frac{\left[(\gamma-3)Z_1^2+(\gamma-3)\Bar{Z_1}^2-4(\gamma+3)Z_1\bar{Z_1}\right]}{32(\gamma-3)(\gamma+1)}
\end{align}
to second order in $Z_1$. As a preliminary check on this result we can test the linear limit. The modulation equations \eqref{eq:weaklynonlinear:modulation_eqns_in_z} may be combined into the oscillator equation
\begin{equation}
    \partial_{TT} A = -2i \left|\frac{\partial ^2 \langle L_{10}\rangle}{\partial T \partial Z_1} \right| B-4 \left| \frac{\partial \left\langle L_{10} \right\rangle}{\partial Z_1}\right|^2 A.
\end{equation}
Inserting our weakly nonlinear averaged Lagrangian and retaining terms at linear order yields oscillatory solutions $A\propto \exp(i\omega_p T)$ with precessional frequency $\omega_p = \pm 1/2$.  This matches onto the linear bending modes with tilting frequency $\omega = 1\pm \epsilon/2$ in the local frame, as found previously in Paper I. Retaining higher order contributions allows us to extend this result for weakly nonlinear forcing warps. More generally, for larger amplitude oscillations we must solve for the forced vertical motions numerically. We will demonstrate this semi-analytical procedure in section \ref{subsection:modulation_precessing_solutions}.

\subsection{Precessing warp solutions}
\label{subsection:modulation_precessing_solutions}

\begin{figure}
    \centering
    \includegraphics[width = \columnwidth]{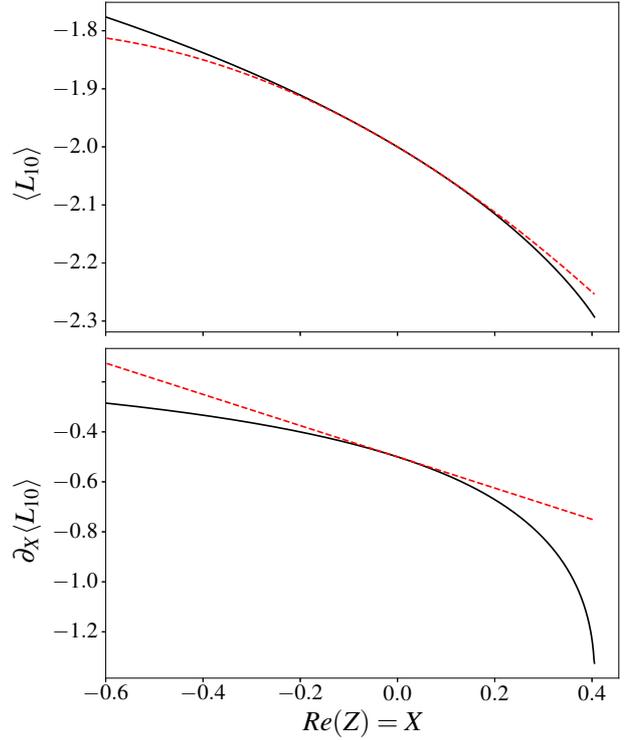}
    \caption{\textit{Upper panel:} $\left\langle L_{10}\right\rangle$ is calculated by averaging equation \eqref{eq:weaklynonlinear:L_10} over the orbital timescale, having numerically identified the periodic solutions for $H$ forced by real $Z_1$ according to equation \eqref{eq:weaklynonlinear:vertical_oscillator}. The gradient of this is then plotted in the lower panel and is related to the precessional frequency as per equation \eqref{eq:modulation:omega_p}. The weakly-nonlinear expansion obtained up to $O(Z_1^2)$ in equation \eqref{eq:weakly_nonlinear_averaged_L} is then over-plotted as dashed red lines, which give the leading order contribution of the nonlinearity. Note the solutions terminate at $X \sim 0.4$ at which point this smooth modulation theory will break down. Meanwhile the solutions may be continued indefinitely for $X<0$.}
    \label{fig:avg_lag}
\end{figure}

With this semi-analytical modulation theory in hand, we will look for a pair of special solutions which correspond to the nonlinear extension of the normal bending modes. In line with the equipartition of tilt and shear found for linear bending waves, we restrict attention to complex amplitudes for which the magnitudes are equal and perfectly in-phase or anti-phased. In this case $B = \pm A$ and thus $Z_1 = \pm |A|^2$ is a real quantity, where the positive/negative sign describes in/anti-phase tilt and shear. Thus we can restrict attention to the averaged Lagrangian along the real line for which we denote $Z_1 = X$. This may be computed numerically using a shooting code which converges to the $2\pi$ periodic solutions for $H$ as shown in Fig.~\ref{fig:avg_lag}. Periodic solutions are found for all $X<0$, which correspond to anti-phased tilt and shear forcing. Meanwhile, the solution terminates in a saddle node bifurcation for sufficiently large $X>0$ (as previously noted by \cite{Ogilvie2013} in the case $\gamma=1$), whereupon this theory breaks down. Observe that the weakly nonlinear solution for $H$, computed to second order in equation \eqref{eq:weakly_nonlinear_averaged_L}, is plotted as the red dashed line and provides a good fit for small $X$.
Making use of Wirtinger complex differentiation,
\begin{equation}
    \frac{\partial}{\partial Z_1} = \frac{1}{2}\left(\frac{\partial }{\partial X}-i\frac{\partial}{\partial Y}\right) 
    \quad \text{and} \quad     
    \frac{\partial}{\partial \bar{Z_1}} = \frac{1}{2}\left(\frac{\partial}{\partial X}+i\frac{\partial}{\partial Y}\right),
\end{equation}
and inserting $B=\pm A$, the amplitude modulation equations become
\begin{equation}
    \label{eq:real_Z_modulation}
    \partial_T A = \pm i\frac{\partial \left\langle L_{10}\right\rangle}{\partial X} A
\end{equation}
when evaluated along the real $Z_1$ line. Note that $Y$ derivatives disappear as $\left\langle L_{10} \right\rangle$ possesses reflectional symmetry about the $X$-axis. This must be the case since the forcing function $f_{ts}$ obtained upon conjugating $Z_1$ is the identical up to a shift in phase. Therefore the periodic solutions for $H$ and hence the averaged Lagrangian must be the same for $Z_1\mapsto\bar{Z_1}$. Examining the form of the modulation equation \eqref{eq:real_Z_modulation} we see it corresponds to a rotation of the complex amplitude whilst the magnitude remains constant. Both oscillator amplitudes rotate at equal rates (since they are described by identical equations) and so the forcing product $Z_1$ remains constant and on the real axis:
\begin{equation}
    \frac{\partial Z_1}{\partial T} = \pm\frac{\partial |A|^2}{\partial T} = \pm \left( A\frac{\partial \bar{A}}{\partial T} + \bar{A}\frac{\partial A}{\partial T}\right) = 0.
\end{equation}
We seek oscillatory solutions of the form $e^{i\omega_p T}$ such that 
\begin{equation}
    \label{eq:modulation:omega_p}
    \omega_p = \pm \frac{\partial \left\langle L_{10} \right\rangle}{\partial X}
\end{equation}
where the $-$ solution corresponds to the anti-phased solutions with $X<0$ and the $+$ solutions correspond to the in phase solutions with $X>0$. Reconstructing the tilting oscillator motion,
\begin{equation}
    J_{31,0} = \sqrt{|X|}\Re[e^{-i(1-\epsilon\omega_p)t}],
\end{equation}
aids the interpretation of the result. The frequency observed in the local model is $\omega = 1-\epsilon\omega_p$. Numerically we see from Fig.~\ref{fig:avg_lag} that $\partial_X \left\langle L_{10} \right\rangle < 0 $ so for the in-phase motions $\omega_p < 0.$ Thus the local frequency is enhanced whilst the period 
\begin{equation}
    \label{eq:smooth_modulation:period}
    T = \frac{2\pi}{1-\epsilon\omega_p}
\end{equation}
is reduced. This recovers the retrograde precession expected for the nonlinear extension of the in-phase bending modes. Similarly, for the anti-phase tilt and shear, $\omega_p>0$ and the oscillation period is less than the orbital period. This may be interpreted as prograde precession of the warped torus structure from a non-rotating global frame.  We will return to these solutions in section \ref{section:shooting_method} where we will verify this theory against the full equation set.

\section{Bouncing regime}
\label{section:bouncing_regime}

Upon reaching a critical warping amplitude, the smooth modulation theory of section \ref{section:smooth_modulation} will break down. Indeed, Fig.~\ref{fig:avg_lag} shows that the averaged Lagrangian solution terminates past a certain forcing amplitude. Furthermore, in section \ref{section:numerical_motivation}, we numerically identified a qualitatively distinct behaviour where the nonlinear vertical oscillator resonantly grows to large amplitudes and becomes extremely compressive. In this section we will develop a separate analytical theory for understanding this regime. We will begin by focusing our attention on the vertical oscillator forced by the warp, which is the defining feature of this bouncing regime, before incorporating the feedback self-consistently onto the tilt and shear.

\subsection{Bouncing vertical oscillator}
\label{section:bouncing_oscillator}
Initially we will ignore the dynamical evolution of the warp, as described by the tilt and shear equations for $J_{31}$ and $J_{13}$ respectively. Instead we treat the warp as being fixed and look for the response of the vertical oscillator as described by equation \eqref{nonlinear_eq: J_33_ode} for $J_{33}$. This approach is similar to that taken by \cite{Ogilvie2013}, to which we uncover a close mathematical correspondence. In order to draw a formal comparison with their analysis, we motivate a coordinate transformation which essentially subtracts the tilting motion and isolates the compressive behaviour. We take $H = J/J_{11}$, which can be interpreted as a measure of the disc thickness since $J$ is proportional to the cross-sectional area and $J_{11}$ approximates the width of the ring. We also fix the the tilt and shear coordinates to oscillate harmonically with some arbitrary phase relationship. Thus, $J_{13} = \Re[A\exp(it)]$ and $J_{31} = \Re[B\exp(it)]$ where we redefine the complex amplitudes $A = a\exp(i\theta_A)$ and $B = b\exp(i\theta_B)$. We will assume that the radial extent of the ring, described by $J_{11}$, is held constant. Indeed, we see in Figs.~\ref{fig:ring_solver_run} and \ref{fig:energy_partition} that the $J_{11}$ oscillator evolves independently from the mode coupling phenomenon so we will ignore equation \eqref{nonlinear_eq: J_11_ode}. Inserting these transformations into equation \eqref{nonlinear_eq: J_33_ode} yields
\begin{equation}
    \label{nonlinear_eq:H_eqn}
    \Ddot{H}+H -\frac{\hat{T}_0}{J_{11}^{\gamma-1}}\frac{1}{H^{\gamma}} = \frac{1}{J_{11}}\Re\left( \frac{3}{2}Z_2 e^{2it}-\frac{1}{2}Z_2\right),
\end{equation}
where the forcing product $Z_2$ is defined as $A\Bar{B}$. Here we are allowing for an arbitrary choice of $J_{11}$ as opposed to the convenient scaling chosen previously in equation \eqref{eq:modulation:J_11_scaling}. This re-scaled definition is simply related to that introduced in our modulation theory by a multiplicative factor, 
\begin{equation}
    \label{nonlinear_eq:Z_2_Z_1}
    Z_2 = J_{11}^2\epsilon Z_1.
\end{equation}
Clearly when $J_{11} = \epsilon^{-1/2}$ as before, we recover the equality between the two definitions. The left-hand side of equation \eqref{nonlinear_eq:H_eqn} represents a free oscillator where the harmonic trajectory is interrupted by the pressure based anharmonic restoring force as the ring is compressed. The right-hand side is a forcing term with a strength proportional to the product of the shear and tilt magnitudes $Z_2$. As we have already seen in section \ref{section:smooth_modulation}, equation \eqref{nonlinear_eq:H_eqn} once again emphasises the generic effect of warped geometries forcing vertical motions. 

\subsubsection{Free vertical oscillation period}
We will now concentrate on the properties of the free non-linear vertical oscillator. To this end we set $Z_2=0$ and work with 
\begin{equation}
    \label{nonlinear_eq: free_oscillator}
    \Ddot{H}+H -\frac{\hat{T}_0}{J_{11}^{\gamma-1}}\frac{1}{H^{\gamma}} = 0.
\end{equation}
This can be derived from a conserved energy Hamiltonian $\mathcal{H}_f$ composed of the sum of kinetic, potential and internal energies respectively
\begin{equation}
    \label{nonlinear_eq: free_oscillator_hamiltonian}
    \mathcal{H}_f = \frac{1}{2}\dot{H}^2+\frac{1}{2}H^2+\frac{\hat{T}_0}{J_{11}^{\gamma-1}}\frac{H^{1-\gamma}}{\gamma-1}.
\end{equation}
Equation \eqref{nonlinear_eq: free_oscillator} clearly permits an equilibrium at $H=\left(\hat{T}_0/J_{11}^{\gamma-1}\right)^{\gamma+1}$ and a linear perturbation then yields a natural oscillation frequency of $\sqrt{\gamma+1}$, as expected from our previous analysis of breathing modes in Paper I. As the amplitude increases into the non-linear regime, we can qualitatively see that the frequency tends monotonically towards $2$. Indeed, in this case $H$ behaves predominantly as a harmonic oscillator in a quadratic potential with an impulsive pressure reversal acting when $H\rightarrow 0$ which rectifies the motion. As the amplitude becomes ever larger, the harmonic motion dominates for the majority of the trajectory. Thus we expect the period of the free oscillator to be $T \sim\pi$ to leading order with a small correction due to the phase shift incurred by pressure. Using the Hamiltonian energy function we can construct an integral for the period as follows:
\begin{equation}
    \label{nonlinear_eq:period_integral}
    T = 2\int_{H_{min}}^{H_{max}}\frac{dH}{\sqrt{2\mathcal{H}_f-\frac{2 \hat{T}_0 H^{1-\gamma}}{J_{11}^{\gamma-1}(\gamma-1)}-H^2}},
\end{equation}
where 
\begin{equation}
    H_{min}\sim \left[\frac{\hat{T}_0}{(\gamma-1)J_{11}^{\gamma-1}\mathcal{H}_f}\right]^{\frac{1}{\gamma-1}}
    \quad \text{and} \quad
    H_{max}\sim \sqrt{2\mathcal{H}_f}  
\end{equation}
denote the minimum and maximum turning points of the vertical oscillator to leading order. Unfortunately this integral cannot be analytically evaluated except in the special case $\gamma=3$, for which we find a period of exactly $\pi$. For other values of $\gamma$ we turn to a range splitting technique which allows us to construct an asymptotic expression in the limit of large amplitude oscillations. This involves approximating the integrand in three distinct intervals and then matching them together such that the errors are subdominant. The leading order deviation from period $\pi$ is found to be
\begin{equation}
\label{nonlinear_eq:asymptotic_period}
    T-\pi = 
    \begin{cases}
          \varpi(\gamma, H_{max}) = c(\gamma)H_{max}^{-1-\gamma}, & \gamma < 2, \\
          \eta(\gamma, H_{min}) = d(\gamma)H_{min}^{(1+\gamma)/2}, & \gamma >2,
    \end{cases}
\end{equation}
with $\gamma$ dependent coefficients
\begin{align}
    & c(\gamma) = \frac{4 \sqrt{\pi} \hat{T}_0}{(\gamma-2)(\gamma-1)J_{11}^{\gamma-1}} \frac{\Gamma(2-\gamma/2)}{\Gamma(1/2-\gamma/2)}, \\
    & d(\gamma) = -\sqrt{\frac{2\pi(\gamma-1)J_{11}^{\gamma-1}}{\hat{T}_0}} \frac{\Gamma(1+1/(1-\gamma))}{\Gamma(1/2+1/(1-\gamma))}.
\end{align}
where $\Gamma$ denote gamma functions. The key point here is that the phase delay has two separate asymptotic limits set by the value of $\gamma$. When $\gamma<2$ the ring is more compressible and the period offset is attributed to the cumulative extended effects of pressure over the trajectory. Meanwhile when $\gamma>2$ the ring is less compressible and the pressure effects are localised near the minimum turning point. For $\gamma < 3$ both $c(\gamma)$ and $d(\gamma)$ are greater than 0 so the period is slightly greater than $\pi$. For the special integrable case $\gamma = 3$, $d(3) = 0$ as expected and the period is exactly $\pi$. For $\gamma$ greater than this, $d(\gamma)$ is negative and the period is slightly less than $\pi$. In the upcoming sections we will assume a typical $\gamma<2$ and hence adopt a period offset $\varpi$ from bounce to bounce.  

\subsubsection{Impulsively forced vertical oscillation}

\begin{figure}
    \centering
    \includegraphics[width=0.8\columnwidth]{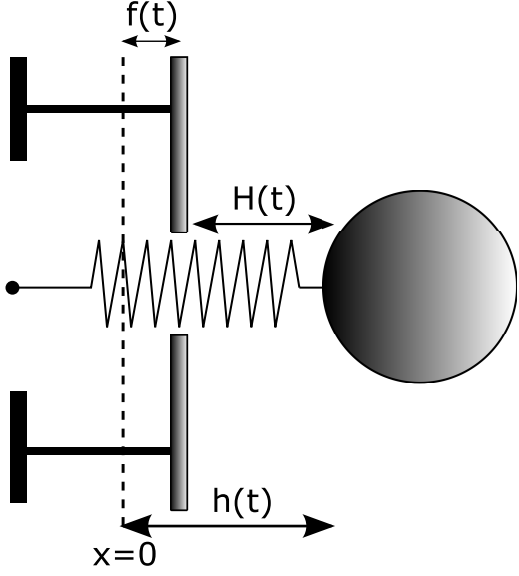}
    \caption{The forced vertical oscillator described by equation \eqref{nonlinear_eq: bouncing_h} may be reinterpreted as a harmonically oscillating mass bouncing off a moving wall. The equilibrium position of the spring-mass system is denoted by the dashed line $x=0$. The wall oscillates about $x=0$ according to $f(t)$ whilst the mass is a distance $H(t)$ from this wall. The spring then has a total extension from its equilibrium $h(t)=f(t)+H(t)$ which incurs a restoring force according to Hooke's law.}
    \label{nonlinear_eq:bounce_cartoon}
\end{figure}

We now extend this analysis to the case where this large amplitude bouncing mode is forced by a fixed warp, with non-zero $Z_2$, as described by equation \eqref{nonlinear_eq:H_eqn}. In fact, we can recast this equation using an intuitive coordinate transformation which reinterprets this forcing term as a localised bouncing off an oscillating boundary. Indeed, if we write
\begin{equation}
    h(t) \equiv H(t) + f(t) = J_{33},
\end{equation}
where $f(t) = J_{13}J_{31}/J_{11}$, and substitute into \eqref{nonlinear_eq:H_eqn} we recover
\begin{equation}
    \label{nonlinear_eq: bouncing_h}
    \Ddot{h}+h = \frac{\hat{T}_0}{J_{11}^{\gamma-1}}H^{-\gamma}.
\end{equation}
This is simply equation \eqref{nonlinear_eq: J_33_ode} in disguise, which may seem a rather circular procedure. However the purpose of introducing this coordinate transformation lies in the helpful physical reinterpretation of the problem. We can view $h(t)$ as the extension of a mass on a spring from an equilibrium position. This wants to undergo harmonic motion according to Hooke's law until the motion is interrupted by an oscillating wall at position $f(t)$. Thus $H(t)$ is the distance between the mass and the wall as visualised in Fig.~\ref{nonlinear_eq:bounce_cartoon}.
When $H(t)$ tends to zero from above the relative velocity between the mass and the wall will reverse in an ideal, elastic bounce. This is very similar to the problem investigated by \citet{Holmes1982} and \citet{Luo1996} with regards to a ball bouncing off an oscillating table, where the motion is reduced to a discrete mapping from bounce to bounce. We proceed similarly by neglecting the pressure contribution from $H^{-\gamma}$ in between bounces. Instead, we assume that it acts impulsively to reverse the direction of motion upon each elastic collision with the wall. Furthermore, it incurs a small phase delay $\varpi(\gamma,H_{max})$, in accordance with our asymptotic investigation of the free vertical oscillator as presented in equation \eqref{nonlinear_eq:asymptotic_period}. These assumptions are valid provided the amplitude and velocity of the mass motion are much larger than the wall position $f$ and velocity $\dot{f}$ at the time of impact. In this case, the bouncing period only slightly departs from $\pi$ and thus the phase relationship with respect to the oscillating wall evolves slowly. Assume that the $n^{th}$ bounce occurs at $t_n = (n-1)\pi+\phi_n$, where $\phi_n$ is a phase offset which evolves slowly. Just after the bounce we have
\begin{equation}
    H_n = 0,  \quad \dot{H}_n \equiv v_n>0,
\end{equation} 
and the wall has position and velocity given by
\begin{align}
    & f_n = \frac{ab}{2 J_{11}}[\cos(\theta_A+\theta_B+2\phi_n)+\cos(\theta_A-\theta_B)], \\ 
    & \dot{f}_n = -\frac{ab}{J_{11}}\sin (\theta_A+\theta_B+2\phi_n).
\end{align}
Thus the position and velocity of the mass are
\begin{equation}
    \label{nonlinear_eq:bounce_condition}
    h_n = f_n, \quad \dot{h}_n = \dot{H}_n+\dot{f}_n = v_n+\dot{f}_n.
\end{equation}
Between bounces we assume purely harmonic motion governed by $\Ddot{h}+h=0$. The initial conditions \eqref{nonlinear_eq:bounce_condition} then determine the trajectory
\begin{equation}
    h = f_n\cos(t-t_n)+(v_n+\dot{f}_n)\sin{(t-t_n)}.
\end{equation}
However, we wish to capture the retarding effect of pressure so we incorporate the phase offset $\varpi$ taken from our asymptotic analysis of the free non-linear vertical oscillator as follows:
\begin{align}
    h &= f_n\cos(t-t_n-\varpi_n)+(v_n+\dot{f}_n)\sin(t-t_n-\varpi_n), \\
    \dot{h} &= -f_n\sin(t-t_n-\varpi_n)+(v_n+\dot{f}_n)\cos(t-t_n-\varpi_n).
\end{align}
The next bounce occurs at $t_{n+1} = n\pi+\phi_{n+1}$. Substituting this into the above expressions allows us to relate successive bounces as
\begin{align}
    \label{nonlinear_eq:f_update}
    f_{n+1} = &-f_n\cos(\phi_{n+1}-\phi_n-\varpi_n) \nonumber\\
            &- (v_n+\dot{f}_n)\sin(\phi_{n+1}-\phi_n-\varpi_n),\\
    \label{nonlinear_eq:v_update}
    v_{n+1} = &-f_n\sin(\phi_{n+1}-\phi_n-\varpi_n) \nonumber\\         &+(v_n+\dot{f}_n)\cos(\phi_{n+1}-\phi_n-\varpi_n)+\dot{f}_{n+1}.
\end{align}
In the large amplitude limit with $v_n \gg |f_n|,|\dot{f}_n|$, the terms in the equation \eqref{nonlinear_eq:f_update} can only be consistently balanced provided $|\phi_{n+1}-\phi_n-\varpi_n|\ll 1$. Since $\varpi$ is a small phase correction, this in turn ensures $|\phi_{n+1}-\phi_n| \ll 1$. This agrees with our expectation that the phase evolves slowly from bounce to bounce. With this assumption, these equations can be simplified to leading order giving the recursive update scheme
\begin{align}
    \label{nonlinear_eq:phi_mapping}
    \phi_{n+1}-\phi_{n} &= \varpi_n-\frac{2f_n}{v_n}, \\
    \label{nonlinear_eq:v_mapping}
    v_{n+1} - v_{n}     &= 2\dot{f}_n. 
\end{align}
The $\phi$ update is composed of two parts -- the $\varpi$ contribution from pressure and also the effect of the oscillating impact position. Recognising that the amplitude of the oscillating mass is approximately equal to the impact velocity with the wall, we can then express the phase delay as $\varpi_n(v_n) = c(\gamma)v_n^{-(\gamma+1)}$ in accordance with \eqref{nonlinear_eq:asymptotic_period}.

\subsubsection{Hamiltonian structure}
\label{subsubsection:vertical_hamiltonian}

\begin{figure}
    \centering
    \includegraphics[width=\columnwidth]{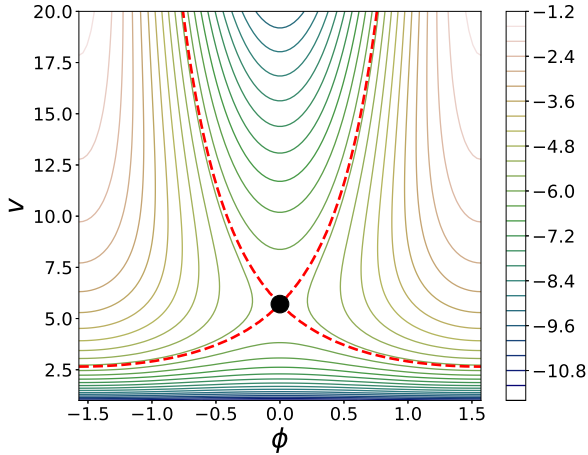}
    \caption{$\phi -v$ phase plane portrait for the specific choice $\theta_A = \theta_B=0$, $\gamma = 5/3$, $J_{11}=100$ and $\hat{T}_0=100^{\gamma-1}$. The coloured solid lines denote contour levels of the Hamiltonian $\mathcal{H}$, along which the system evolves from bounce to bounce. The red dashed lines denote the separatrix curves originating from the critical unstable saddle points and delimit the resonant, phase locking behaviour from the periodic behaviour.}
    \label{nonlinear_eq:hamiltonian_phase_plane}
\end{figure}

Considering the variable updates from bounce to bounce are small, we may take the continuous ODE analogue of these discrete mappings to be
\begin{align}
    & \frac{d\phi}{dn} = \varpi - \frac{2 f}{v}, \\
    & \frac{d v}{d n} = 2\dot{f}.
\end{align}
As we might anticipate for an ideal system, these equations possess an autonomous symplectic structure. This is best seen by the change of variables $I = \frac{1}{2}v^2$, which is the classical action of a harmonic oscillator. The Hamiltonian is then found to be
\begin{equation}
    \label{nonlinear_eq:pendulum_hamiltonian}
    \mathcal{H} 
    = \frac{c(\gamma)}{(1-\gamma)}(2I)^{(1-\gamma)/2}-2(2I)^{1/2}f(\phi),
\end{equation}
with the canonical equations of motion
\begin{equation}
    \frac{d\phi}{dn} = \frac{\partial \mathcal{H}}{\partial I}, \quad \frac{d I}{dn} = -\frac{\partial \mathcal{H}}{\partial \phi}.
\end{equation}
Note that when the warped forcing is absent, the Hamiltonian is independent of the phase angle and hence the action is invariant whilst the phase advances uniformly. This simply corresponds to the free harmonic oscillator with constant amplitude and phase delay from bounce to bounce. More generally for non-zero forcing, the contours of the Hamiltonian trace out the trajectories in phase space. An example of this structure is shown in Fig.~\ref{nonlinear_eq:hamiltonian_phase_plane} for the particular choice $\theta_A = \theta_B = 0$, which corresponds to the tilt and shear oscillators being in-phase. Here we set the value of $c(\gamma;J_{11},\hat{T}_0)$ for $\gamma=5/3$, which also depends on the scaling parameters chosen for the ellipse. As per our numerical runs in section \ref{section:numerical_motivation} we choose $J_{11}=100$ and the value of $\hat{T}_0$ so the associated equilibrium ring has aspect ratio $\epsilon=0.01$. Note that the structure is $\pi$ periodic since the phase variable $\phi$ is measured modulo the rectified harmonic period of $\pi$. The red dashed line plots the hetero-clinic separatrix structure emanating from the unstable saddle point located at 
\begin{equation}
    \label{eq:bouncing:forced_vertical_saddle}
    \{0,[c(\gamma)J_{11}/(2ab)]^{1/\gamma}\}.
\end{equation}
This delimits a circulating solution from a resonantly growing solution which becomes phase locked as the bounce amplitude tends to infinity.
This resonant phase locking is observed in numerical solutions of equation \eqref{nonlinear_eq:H_eqn} and helps elucidate the physical mechanism responsible for the growth of compressive vertical motions. When the phase delay incurred by the pressure retardation is sufficiently counteracted by the changing phase relationship with the wall, energy is constructively input into the breathing mode over many cycles. This increases its amplitude and reduces the rate of future phase evolution, further locking it into a resonant relationship. This distinct behaviour for sufficiently large warps hints towards the existence of a critical warping amplitude above which the oscillator will be driven into the bouncing regime as we found in our numerical experiments in section \ref{section:numerical_motivation}. 

\subsubsection{Periodic structure using shooting method}
\label{subsubsection:forced_vertical_shoot}
%
\begin{figure}
    \centering
    \includegraphics[width=\columnwidth]{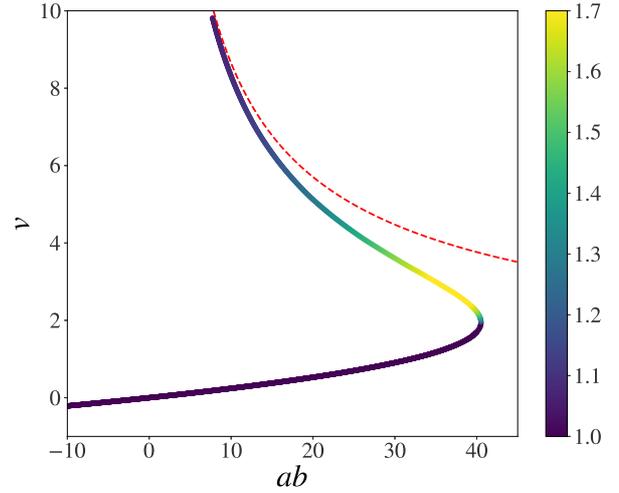}
    \caption{The thick line tracks periodic solutions of equation  \eqref{nonlinear_eq:H_eqn} with the phase variables set to be $\theta_A = \theta_B$ and for $\gamma = 5/3$, $J_{11}=100$ and $\hat{T}_0=100^{\gamma-1}$. The x-axis plots the value of $a b$ which is a measure of the tilt and shear forcing amplitude. The y-axis plots the maximum value $v = \dot{H}$ for each periodic solution. The branch colour denotes the maximum magnitude of the eigenvalues obtained from the monodromy matrix associated with a Floquet analysis about each periodic solution. Values above 1 indicate instability. The dashed red line denotes the analytical prediction for the location of the saddle point in the large bounce regime according to equation \eqref{eq:bouncing:forced_vertical_saddle}.}
    \label{fig:forced_vertical_periodic}
\end{figure}
The analytical progress made in the previous section accurately describes the forced vertical oscillator in the extreme bouncing regime. However, in the case of low amplitude oscillations, not far from the equilibrium of the disc, we might expect our approximations to break down. Indeed, previous work by \cite{Ogilvie2013} found stable periodic solutions for the simple laminar flows in a warped disc, provided the enforced warp amplitude is sufficiently low. In order match the high and low amplitude regimes, we use a shooting scheme to identify the existence of periodic solutions as the imposed warp is varied through the tilt and shear product $ab$. The shooting code implemented solves equation \eqref{nonlinear_eq:H_eqn}, with $\gamma=5/3$, $J_{11}=100$ and $\hat{T}_0$ such that the equilibrium ring has aspect ratio $\epsilon=0.01$. We use a typical Runge-Kutta integrator with adaptive step-size and then minimise residuals at the boundary in accordance with Levenberg–Marquardt least squares optimisation \citep{Dednam2014}. 

This efficiently converges onto $2\pi$ periodic solutions which are plotted in Fig.~\ref{fig:forced_vertical_periodic}. The tilt and shear oscillators are set with the phase relationship $\theta_A = \theta_B$. The y-axis denotes the maximum value of $v = \dot{H}$ which is the appropriate amplitude measure for the periodic solutions. Meanwhile, the x-axis describes the forcing product $ab$. When $ab > 0$ the tilt and shear are in phase, whilst when $ab < 0 $ the tilt and shear are in anti-phase. For each periodic solution we also perform a Floquet stability analysis. We calculate the monodromy matrix and extract the eigenvalues with maximum magnitude. Since we are expanding about a periodic solution, there always exists an eigenvalue equal to 1 which corresponds to a perturbation tangential to the periodic solution. If there exists an eigenvalue with absolute magnitude greater than 1 (i.e. outwith the complex unit circle), the periodic solution is unstable. The value of the periodic solutions are coloured according to the maximum magnitude eigenvalue, with purple denoting the stable solution baseline with eigenvalue equal to 1. Finally, the saddle point location, predicted by the high amplitude bouncing theory, is plotted as the red dashed line for comparison.

In the high amplitude limit, we do indeed converge to the unstable saddle-point solutions with an in-phase forcing $ab>0$. For large values of $v$ the analytically predicted solution tends asymptotically towards our numerical findings. As we move down this branch towards the kink, the numerical shooting code deviates from our prediction as the impulsive approximation breaks down. The saddle point exhibits a peak unstable growth rate before the kink turns over and enters the stable lower branch. This can be continued indefinitely towards large negative values of $ab$. When $ab$ becomes less than 0, this is equivalent to the tilt and shear becoming $\pi$ out of phase. This in turn changes the phase relationship with the driven vertical oscillator so the velocity amplitude becomes negative. Whilst the upper branch describes a saddle point, the numerically identified lower branch represents a stable centre. As the forcing warp is increased towards the turning point, these two points converge and eventually collide in a saddle node bifurcation at $ab\sim 40$. This behaviour is consistent with the termination of solutions found previously for the forced vertical oscillator within the context of our modulation theory in section \ref{subsection:forced_vertical}. In Fig.~\ref{fig:avg_lag} the solution branch ends abruptly at $Z_1 = 0.4$. When this is re-scaled by $J_{11}^2\epsilon=100$ to account for our arbitrary choice of ring width we find agreement with $Z_2 = a b =40$. This bifurcation point sets a critical warping amplitude beyond which no periodic solutions can be found for in phase tilt and shear. Instead, trajectories are carried up along the steep contours as shown in Fig.~\ref{nonlinear_eq:hamiltonian_phase_plane} tending asymptotically to the fixed resonant phase relationship $\phi = \pm \pi/2$. 

These results agree with the findings of \cite{Ogilvie2013}. They also find that for a sufficiently large positive warped forcing, the $\pi$ periodic solutions terminate. This offers a mechanism for which a system with no initial vertical motion can be driven to large amplitudes, provided the warp amplitude lies beyond this critical turning point. This is what we see in Fig.~\ref{fig:ring_solver_run} where, for moderate warp, the the vertical motion becomes highly activated. As growth continues, the feedback of the vertical motion onto the warp will become important and the enforced warp assumption will also break down. We will address this via a self-consistent coupling of the warp to the vertical bouncing in the next section. 

\subsection{Feedback onto the warp}
\label{subsection:coupling_warp}

The previous analysis assumes that the warp is fixed, with the tilt and shear oscillating sinusoidally at the orbital frequency. We found that this leads to resonant growth if the phase becomes locked and energy continues to be injected into the bouncing motions. In reality, total energy is conserved and energy flowing into one mode must be coupled with energy leaving another, as seen in the motivating plots of Fig.~\ref{fig:energy_partition}. Indeed, we must consider the back-reaction onto the warp which is then allowed to evolve.

Let us consider the case that the breathing mode has entered into the highly compressive non-linear regime. We have seen that the effect of pressure can be treated as an impulsive forcing which reverses the direction of the bouncing mass. This also gives us reason to believe that the pressure terms in the equations \eqref{nonlinear_eq: J_13_ode} and \eqref{nonlinear_eq: J_31_ode} also enter as time localised impulsive forces. Indeed, in this regime we expect the tilt and shear oscillators to undergo linear harmonic motion which is periodically kicked, causing an instantaneous change in their amplitude and phase from bounce to bounce. By connecting the piece-wise harmonic intervals between the $n^{th}$ and $(n+1)^{th}$ bounces, we can create an iterable mapping for the evolution of the system. We see from equation \eqref{nonlinear_eq: bouncing_h} that the impulsive forcing on the right hand side $\hat{T}_0/J_{11}^{\gamma-1}H^{\gamma}$ provides this Dirac delta forcing. It reverses the impact velocity $v_{n+1}$ of the mass relative to the wall such that
\begin{equation}
\label{nonlinear_eq:dirac_delta_forcing}
\frac{\hat{T}_0}{J_{11}^{\gamma-1}} \frac{1}{H^{\gamma}} = \frac{\hat{T}_0 J_{11}}{J^{\gamma}} =2v_{n+1}\delta(t-t_{n+1}),
\end{equation}
where $\delta(t)$ is the Dirac delta function. Thus our tilt and shear oscillator equations have the form
\begin{align}
    \label{nonlinear_eq:impulse_forced_J13_J31}
    \Ddot{J}_{13}+J_{13} &=  -\frac{2v_{n+1}J_{31}}{J_{11}}\delta(t-t_{n+1}), \\
    \Ddot{J}_{31}+J_{31} &= 
    - \frac{2v_{n+1} J_{13}}{J_{11}}\delta(t-t_{n+1}).
\end{align}
where we have neglected (the often small constant) $C_x$. These have the general form of harmonic oscillators undergoing impulsive kicks as described by the equation
\begin{equation}
\label{nonlinear_eq:kicked_harmonic_oscillator}
    \Ddot{x}(t)+\omega^2 x(t) = F_{0}\delta(t-t_{n+1}),
\end{equation}
where $F_0$ is the momentum impulse, such that integration over the equation gives an instantaneous change in velocity $\Delta\Dot{x} = F_0$. This problem is completed by furnishing it with the initial conditions $x(0)=x_{0}$ and $\dot{x}(0)=v_{0}$. This equation has wide reaching physical applications and has been studied extensively with application to both classical and quantum problems. The solution is easily found by converting it to an algebraic equation via the Laplace transform and then inverting back to the original variable domain. We find the solution to be
\begin{equation}
    \label{nonlinear_eq_eq:kicked_oscillator_solution}
    x(t) = x_{0}\cos(\omega t)+\frac{v_0}{\omega}\sin(\omega t)+\frac{F_0}{\omega}u(t-t_{n+1})\sin[\omega(t-t_{n+1})],
\end{equation}
where $u$ is the unit-step function. Clearly the amplitude and phase of the oscillator are modified after the impact. We will find it convenient to describe this in terms of a complex amplitude $\chi=\chi_R+i\chi_I = |\chi|\exp(i\theta_{\chi})$ such that
\begin{align}
    x(t) &= \Re[\chi e^{i\omega t}]\\
         &= \frac{1}{2}\left(\chi e^{i\omega t}+\Bar{\chi}e^{-i\omega t} \right) \\
         &= \chi_R\cos(\omega t)-\chi_I\sin(\omega t) \\ 
         &= |\chi|\cos(\omega t +\theta_\chi) 
\end{align}
By comparing the sine and cosine coefficients before and after the bounce we find the complex amplitude mapping
\begin{align}
    \chi_{R,n+1} &= \chi_{R,n}-\frac{F_0}{\omega}\sin(\omega t_{n+1}), \\
    \chi_{I,n+1} &= \chi_{I,n}-\frac{F_0}{\omega}\cos(\omega t_{n+1}),
\end{align}
which is equivalent to
\begin{equation}
    \label{nonlinear_eq: complex_mapping}
    \chi_{n+1} = \chi_{n}+\frac{F_0}{\omega}e^{-i(\omega t_{n+1}+\pi/2)}.
\end{equation}
We now apply this method to the equations for $J_{13}$ and $J_{31}$. Inserting the relevant Dirac-delta forcing coefficients leads to the iterative scheme
\begin{align}
\label{nonlinear_eq:lambda_update}
A_{n+1} &= A_{n}-\frac{2 v_n}{J_{11}}\Re(B_n e^{i t_{n+1}}) e^{-i(t_{n+1}+\pi/2)}, \\
\label{nonlinear_eq:chi_update}
B_{n+1} &= B_{n}-\frac{2 v_n}{J_{11}}\Re(A_n e^{i t_{n+1}}) e^{-i(t_{n+1}+\pi/2)}.
\end{align}
The forcing coefficient is proportional to $v_n$ which is set by the bouncing vertical oscillator. Thus in order to close this discrete set we must couple it to the mappings of $v_n$ and $\phi_n$ derived previously in equations \eqref{nonlinear_eq:phi_mapping} and \eqref{nonlinear_eq:v_mapping}. As the tilt and shear evolve from bounce to bounce, this in turn modifies the forcing on the breathing motions in accordance with
\begin{align}
    f(t_n) &= \frac{a_n b_n}{J_{11}}\cos( t_n+\theta_A)\cos(t_n+\theta_B).
\end{align}

\section{Resonant centres for coupled system}
\label{section:resonant_centres}

Having developed an impulsive theory for the bouncing regime, we will now seek special resonant solutions for which the amplitude of the oscillators is constant and the phase relationship between them remains fixed. Physically, these describe large amplitude, globally precessing warped solutions with extreme compressions twice per orbit. 

\subsection{Hamiltonian structure and resonant centres}

We have derived a self-consistent set of discrete equations mapping the non-linear breathing and warping motions at each compression of the ring. This is still highly coupled and requires some further simplification to gain more dynamical insight. We again take the continuous limit (as done previously for the non-linear vertical oscillator) and study the resulting system of ODEs. By writing $A = ae^{i\theta_A}$ and $B = be^{i \theta_B}$ we can split up the real and imaginary parts of \eqref{nonlinear_eq:lambda_update} and \eqref{nonlinear_eq:chi_update} to generate evolutionary equations for the amplitudes and phases:
\begin{align}
\frac{da}{dn} &= \frac{v b}{J_{11}}\left[\sin(\theta_A-\theta_B)+\sin(\theta_A+\theta_B+2\phi)\right], \\
\frac{db}{dn} &= \frac{v a}{J_{11}}\left[-\sin(\theta_A-\theta_B)+\sin(\theta_A+\theta_B+2\phi)\right], \\
a\frac{d\theta_A}{dn} &= \frac{v b}{J_{11}}\left[\cos(\theta_A-\theta_B)+\cos(\theta_A+\theta_B+2\phi)\right], \\
b\frac{d\theta_B}{dn} &= \frac{v a}{J_{11}}\left[\cos(\theta_A-\theta_B)+\cos(\theta_A+\theta_B+2\phi)\right].
\end{align}
Since the motions of all three oscillators are essentially harmonic between bounces, we anticipate that simple \textit{action-angle} coordinates will further elucidate the structure of our equation set. We naturally adopt $\theta_A$, $\theta_B$ and $\theta_C \equiv -\phi$ as our angles whilst the energy of each oscillator gives the actions $I_A = a^2/2$, $I_B = b^2/2$ and $I_C = v^2/2$. Note that taking the negative of $\phi$ makes sense as an angle since an increase in $\phi$ represents a delay to the bounce time. This corresponds to a negative shift in the phase angle of a rectified harmonic oscillator. Using these transformations leads to a set of 6 equations:
\begin{align}
    \label{nonlinear_eq:theta_1}
    & \frac{d\theta_A}{dn}  = \frac{\sqrt{2}}{J_{11}}\sqrt{\frac{I_B I_C}{I_A}}\left[ \cos(\theta_A-\theta_B)+\cos(\theta_A+\theta_B-2\theta_C)\right], \\
    \label{nonlinear_eq:I_A}
    & \frac{d I_A}{dn}  = \frac{2\sqrt{2}}{J_{11}}\sqrt{I_A I_B I_C}\left[ \sin(\theta_A-\theta_B)+\sin(\theta_A+\theta_B-2\theta_C)\right], \\
    \label{nonlinear_eq:theta_2}
    & \frac{d\theta_B}{dn}  = \frac{\sqrt{2}}{J_{11}}\sqrt{\frac{I_A I_C}{I_B}}\left[ \cos(\theta_A-\theta_B)+\cos(\theta_A+\theta_B-2\theta_C)\right], \\
    \label{nonlinear_eq:I_B}
    & \frac{d I_B}{dn}  = \frac{2\sqrt{2}}{J_{11}}\sqrt{I_A I_B I_C}\left[ -\sin(\theta_A-\theta_B)+\sin(\theta_A+\theta_B-2\theta_C)\right], \\
    \label{nonlinear_eq:theta_3}
    & \frac{d\theta_C}{dn}  = \frac{\sqrt{2}}{J_{11}}\sqrt{\frac{I_A I_B}{I_C}}\left[ \cos(\theta_A-\theta_B)+\cos(\theta_A+\theta_B-2\theta_C)\right] -\varpi, \\
    \label{nonlinear_eq:I_C}
    & \frac{d I_C}{dn}  = -\frac{4\sqrt{2}}{J_{11}}\sqrt{I_A I_B I_C}\sin(\theta_A+\theta_B-2\theta_C).
\end{align}
These possess a symplectic structure amenable to a Hamiltonian formalism. The appropriate Hamiltonian is found to be
\begin{equation}
    \label{nonlinear_eq:coupled_H}
    \mathcal{H} = \frac{2\sqrt{2}}{J_{11}}\sqrt{I_A I_B I_C}[\cos(\theta_A-\theta_B)+\cos(\theta_A+\theta_B-2\theta_C)]+\Delta(I_C)
\end{equation}
where $\Delta(I_C)$ is defined such that $d\Delta/dI_C=-\varpi(\gamma,I_C)$ and characterises the retarding phase offset from the vertical oscillator. Hamilton's equations are then given by
\begin{equation}
    \frac{d \theta_i}{dn} = \frac{\partial H}{\partial I_i}, \quad \frac{d I_i}{dn} = -\frac{\partial H}{\partial \theta_i}.
\end{equation}
This nicely extends the Hamiltonian structure for the forced vertical oscillator found previously in section \ref{subsubsection:vertical_hamiltonian}, which is recovered by fixing the action and angle variables corresponding to $J_{13}$ and $J_{31}$. Notice also the inherent symmetry in the Hamiltonian upon a constant translation in the angles $\theta_i \rightarrow \theta_i+\delta\theta$. This canonical transformation is facilitated by the arbitrariness of setting the phase origin and, via Noether's theorem, is generated by the conserved total action $I_t = I_A+I_B+I_C$. This is reminiscent of the energy conservation as seen in Fig.~\ref{fig:energy_partition} where now the action variables are a proxy for the energy contained in the different modes. The dynamical evolution in phase space allows for an action interchange between modes but constrains trajectories to lie on contours of conserved total action.  

Phase locking now occurs if the resonant angle combinations $\xi_{1} \equiv \theta_A -\theta_B$ and $\xi_{2} \equiv \theta_A+\theta_B-2\theta_C$ are librating around a fixed centre rather than circulating. These fixed points are located by solving $\dot{\xi}_1=\dot{\xi}_2=\dot{I}_i=0$. Resonance requires that $\xi_1 = l\pi$ and $\xi_2 = m \pi$ where $l,m \in \mathbb{Z}$. The freedom to choose the phase relationship permits two separate resonant centres; an upper and a lower branch corresponding to the plus and minus sign respectively in $\cos\xi_1+\cos\xi_2=\pm 2$. The upper branch requires $\theta_A=\theta_B$ so the tilt and shear are in phase. Meanwhile the lower branch requires that $\theta_A=\theta_B+\pi$ so the tilt and shear are exactly out of phase. $\theta_C$ is either $0$ or $\pi$ such that the bounce point of the compressive breathing mode coincides with the points of maximal tilt and shear. This allows us to solve for the action centres
\begin{equation}
    \label{nonlinear_eq:resonant_centres}
    I_{0}(I_C) \equiv I_A = I_B = I_C\pm\frac{\varpi(I_C)\sqrt{I_C} J_{11}}{2\sqrt{2}},
\end{equation}
where the common function $I_0$ equals the equipartition between $I_A$ and $I_B$. Again here, the plus family of solutions correspond to the in-phase tilt and shear, whilst the minus branch correspond to the out of phase solution. We see a continuous family of resonant centres parameterised by the breathing action $I_C$. 
\begin{figure}
    \centering
    \includegraphics[width=\columnwidth]{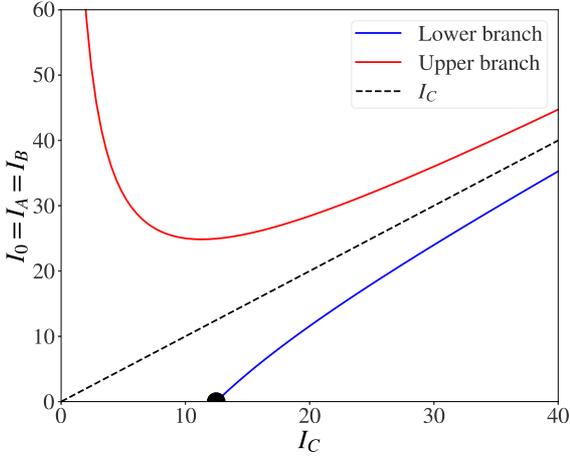}
    \caption{A plot of the upper (red line) and lower (blue line) resonant branches corresponding to the centres identified in equation \eqref{nonlinear_eq:resonant_centres}. The bifurcation points in the dynamical behaviour are noted with black dots and the asymptotic limit of $I_0 = I_C$ is plotted as a dashed black line.}
    \label{fig:resonant_branches}
\end{figure}
%

\subsection{Exploring the branch structure}
\label{subsection:branch_structure}
This branch structure is plotted in Fig.~\ref{fig:resonant_branches} for $\gamma = 5/3$, $J_{11} = 100$ and $\hat{T}_0 = J_{11}^{\gamma-1}$. 
The $x-$axis corresponds to the purely vertical breathing modes with only
the vertical $I_C$ action excited. Numerical evaluation of Hamilton’s
equations \eqref{nonlinear_eq:theta_1} -- \eqref{nonlinear_eq:I_C} suggests that this is a stable periodic solution until reaching a pitchfork bifurcation at the point
\begin{equation}
    \label{nolinear:bifurcation_point}
    I_{3,bif} = \left[\frac{c(\gamma)J_{11}}{4}2^{-\gamma/2}\right]^{\frac{2}{\gamma+2}},
\end{equation}
where the lower blue branch intercepts the $I_C$ axis. At this point, the breathing mode becomes unstable to warping motions and a stable mixed-mode branch is spawned. The tilt and shear are out of phase and their action is slightly less than that of the vertical oscillator as indicated by the dashed line in  Fig.~\ref{fig:resonant_branches}. Formally this bifurcation point arises as a parametric instability of the tilt and shear oscillators which we will now demonstrate. Akin to the classic example of a swing being pumped at twice its natural frequency, the breathing mode pumps the tilt and shear as we increase its amplitude and the frequency becomes sufficiently close to $2$.  Diagonalising equations \eqref{nonlinear_eq: J_13_ode} and \eqref{nonlinear_eq: J_31_ode} we have
\begin{equation}
\ddot{q}_1+\left(1+\frac{\hat{T}_0}{J_{11}^\gamma H^\gamma}\right)q_1 = 0 \quad \text{and} \quad
\ddot{q}_2+\left(1-\frac{\hat{T}_0}{J_{11}^\gamma H^\gamma}\right)q_2 = 0,
\end{equation}
where we have used the change of basis
\begin{equation}
    \begin{pmatrix}
    J_{13} \\
    J_{31}
    \end{pmatrix} = 
    \begin{pmatrix}
    1 &  1 \\
    1 & -1
    \end{pmatrix}
    \begin{pmatrix}
    q_1 \\
    q_2
    \end{pmatrix}.
\end{equation}
Thus the $q_1$ component corresponds to the in-phase tilt and shear contribution and $q_2$ the anti-phase component. Since it is the lower, anti-phased branch which is spawned from the vertical breathing mode x-axis in  Fig.~\ref{fig:resonant_branches}, we proceed to look for parametric instability in the $q_2$ equation. As before, we treat the forcing by the large amplitude breathing mode impulsively so $\hat{T}_0/\left( J_{11} H\right)^{\gamma} = 2 v_{imp}\delta(t-t_{imp})/J_{11}$ where $v_{imp}$ and $t_{imp}$ denote the impact velocity and time respectively. Then we can write the impulsive system of differential equations as
\begin{equation}
    \dot{
    \begin{pmatrix}
    q \\
    p
    \end{pmatrix}
    }
    =\begin{pmatrix}
    0 & 1 \\
    -1 & 0 
    \end{pmatrix}
    \begin{pmatrix}
    q \\
    p
    \end{pmatrix}
    \quad \text{for} \quad t \neq t_{imp},
\end{equation}
\begin{equation}
    \begin{pmatrix}
    \Delta q \\
    \Delta p
    \end{pmatrix}
    =\begin{pmatrix}
    0 & 0 \\
    \frac{2 v_{imp}}{J_{11}} & 0 
    \end{pmatrix}
    \begin{pmatrix}
    q \\
    p
    \end{pmatrix}
    \quad \text{for} \quad t = t_{imp},
\end{equation}
where $q = q_2$ and $p = \dot{q_2}$. $\Delta p$ and $\Delta q$ denote the discrete jump in the quantities at the impact times of the breathing mode $t_{imp}$. This form is now amenable to the impulsive Floquet theory developed by \cite{Bainov1993}. Similar to the usual continuous Floquet analysis, we construct the monodromy matrix $M$ which captures the evolution of the system over one bounce period $T$,
\begin{equation}
    \label{nonlinear_eq:modronomy}
    M = 
    \begin{pmatrix}
    \cos T & \sin T \\
    \frac{2 v_{imp}}{J_{11}}\cos T -\sin T & \frac{2 v_{imp}}{J_{11}}\sin T + \cos T 
    \end{pmatrix}.
\end{equation}
The eigenvalues of $M$ correspond to Floquet multipliers $\mu_i$ which determine the stability of the trivial solution $(q,p)=(0,0)$. Stability requires that $|\mu_i|\leq 1$ for all $i$. It should also be noted from the determinant of $M$ that $\mu_1\mu_2=1$. The characteristic equation for $M$ gives 
\begin{equation}
    \mu_{1,2} = \cos{T}+\frac{v_{imp}}{J_{11}}\sin{T} \pm \sqrt{(\cos{T}+\frac{v_{imp}}{J_{11}}\sin{T})^2-1}.
\end{equation}
If $|\cos{T}+\frac{v_{imp}}{J_{11}}\sin{T}|<1$ then the Floquet multipliers are complex conjugates. Since $\mu_1\mu_2=1$ this requires $|\mu_i|=1$ and the trivial solution is stable. In contrast, if $|\cos{T}+\frac{v_{imp}}{J_{11}}\sin{T}|>1$ then the multipliers are real and distinct. Therefore one must be greater than 1 and the trivial solution is unstable to parametric growth. Let us insert the bouncing period $T = \pi+\varpi$ into this criterion such that
\begin{equation}
    \cos{\varpi}+\frac{v_{imp}}{J_{11}}\sin{\varpi}>1.
\end{equation}
Expanding terms to $\mathcal{O}(\varpi^2)$ in the small phase delay allows us to deduce the instability criterion $v_{imp}/J_{11}-\varpi/2 > 0$. As before, the phase delay can be written asymptotically in accordance with equation \eqref{nonlinear_eq:asymptotic_period} as $\varpi = c(\gamma)v_{imp}^{-1-\gamma}$, so rearrangement yields the critical value
\begin{equation}
    \label{nonlinear_eq:parametric_bifurcation_point}
    v_{imp,bif} = \left[ \frac{c(\gamma)J_{11}}{2}\right] ^\frac{1}{2+\gamma}. 
\end{equation}
This agrees exactly with the bifurcation point identified as the x-intercept of the lower resonant branch in equation \eqref{nolinear:bifurcation_point} and demonstrates the underlying parametric mechanism.

\begin{figure}
    \centering
    \includegraphics[width=\columnwidth]{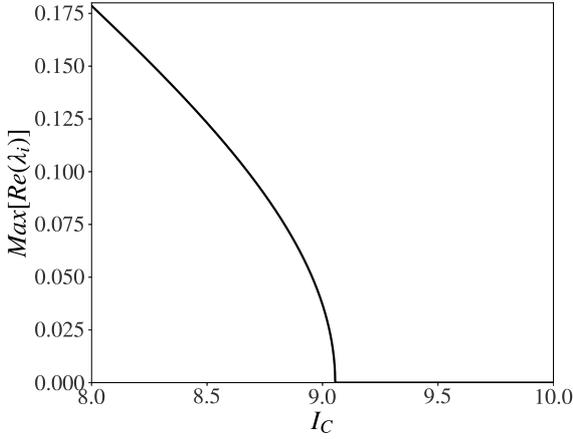}
    \caption{The maximum real eigenvalue for the Jacobian of our action-angle equations evaluated about the upper resonant branch for the case $J_{11}=100$ and $\gamma=5/3$. This is parameterised by the vertical mode action $I_C$. We see instability for $I_C<9.05$ and stability for $I_C>9.05$.}
    \label{fig:nonlinear:upper_branch_stability}
\end{figure}

We are also able to deduce the stability of the non-trivial resonant branches themselves. Consider the equations for $\xi_1$, $\xi_2$, $I_A$, $I_B$ and $I_C$. Linearising about the fixed resonant solutions, parameterised by $I_C$, yields a $5\times 5$ Jacobian matrix which encapsulates the stability as we move along the branches. The solution is stable provided no eigenvalues have a real component. We find that the lower branch is stable for all $I_C$ above the parametric bifurcation point from whence it originates. Meanwhile the upper branch shows a transition from an unstable to a stable region. In Fig.~\ref{fig:nonlinear:upper_branch_stability} we plot the maximum real part of the eigenvalues $\lambda_i$, corresponding to the Jacobian computed about the upper branch. We see that the branch is unstable when $I_C< 9.05$ which corresponds to the region left of the black dot as plotted on the upper branch of  Fig.~\ref{fig:resonant_branches}. 
Moreover we find that the eigenvectors associated with the unstable growth correspond to equal in-phase perturbations of tilt and shear. i.e. those which maintain $I_A=I_B$ and $\theta_A = \theta_B$. We will make use of this fact and restrict our attention to the equal amplitude in-phase tilt and shear. This reduces our system of equations to
\begin{align}
    \frac{d\theta_A}{dn} &= \frac{\sqrt{2}}{J_{11}}\sqrt{I_C}\left[1+\cos{(2\theta_A-2\theta_C)}\right], \\
    \frac{d I_{A+B}}{dn} &= \frac{2\sqrt{2}}{J_{11}}I_{A+B}\sqrt{I_C}\sin(2\theta_A-2\theta_C), \\
    \frac{d\theta_C}{dn} &= \frac{\sqrt{2}}{2 J_{11}}\frac{I_{A+B}}{\sqrt{I_C}}\left[1+\cos{(2\theta_A-2\theta_C)}\right]-\varpi(I_C), \\
    \frac{d I_C}{dn} &= -\frac{2\sqrt{2}}{J_{11}}I_{A+B}\sqrt{I_C}\sin(2\theta_A-2\theta_C) ,
\end{align}
where $I_{A+B} = I_A+I_B$ and $I_t = I_{A+B}+I_C$ is still clearly a conserved quantity. This can be derived from the reduced Hamiltonian 
\begin{equation}
   \mathcal{H}= \frac{\sqrt{2}}{J_{11}}I_{A+B}\sqrt{I_C}\left[ 1+\cos{(2\theta_A-2\theta_C)}\right]+\Delta(I_C).
\end{equation}
We can simplify this if there exists a canonical transformation which invokes the conserved total action as one of our momenta. Indeed, the point transformation $\phi_1 = \theta_A-\theta_C$ and $\phi_2=\theta_C$ with conjugate momenta $P_1 = I_{A+B}$ and $P_2 = I_{A+B}+I_C=I_t$ yields the simplified Hamiltonian
\begin{align}
    \label{nonlinear_eq:hamiltonian_reduced_upper}
   \mathcal{H}  &= \frac{\sqrt{2}}{J_{11}}P_1\sqrt{P_2-P_1}\left[1+\cos{(2\phi_1)}\right]+\Delta(P_2-P_1)  \nonumber\\ 
        &= \frac{\sqrt{2}}{J_{11}}I_{A+B}\sqrt{I_t-I_{A+B}}\left[1+\cos{(2\phi_1)}\right]+\Delta(I_t-I_{A+B}).
\end{align}
\begin{figure*}
    \centering
    \includegraphics[width = 0.95\textwidth]{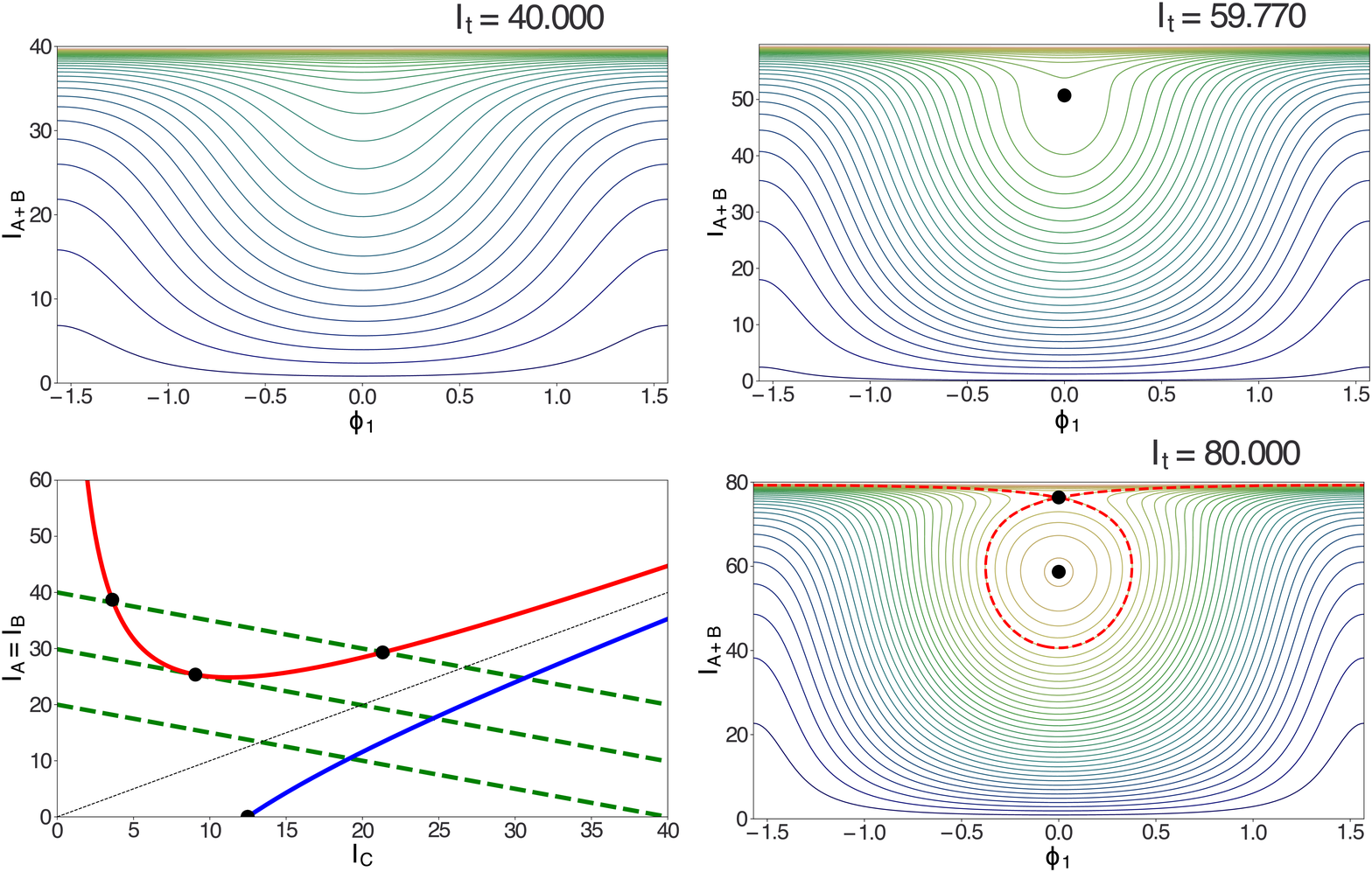}
    \caption{\textit{Top Left}: The resonant branch structure as per Fig.~\ref{fig:resonant_branches} is overlaid with green dashed lines corresponding to $I_{A+B}=I_t-I_C$ for three choices of $I_t = \{40.0,59.77,80.0\}$. Positions where the green dashed line intersects the upper branch result in fixed points, denoted by black dots. The remaining panels show contours of the Hamiltonian \eqref{nonlinear_eq:hamiltonian_reduced_upper} for the different values of $I_t$. We see that as $I_t$ is increased and the green line crosses the upper branch, a saddle-node bifurcation spawns a centre and saddle-point, plotted as black dots in the bottom panels.}
    \label{fig:nonlinear:reduced_upper_portrait}
\end{figure*}
Clearly the absence of $\phi_2$ ensures $I_t$ is conserved. Now we can examine slices in phase space for a choice of constant $I_t$ and visualise the reduced two dimensional structure. Here trajectories of $I_{A+B}$ and $\phi_1$ are traced out by contours of the Hamiltonian. Examining the evolution of the phase portrait as we vary the constant $I_t$, helps us gain further insight into the upper resonant branch. In Fig.~\ref{fig:nonlinear:reduced_upper_portrait} we plot the phase portrait for three different values of $I_t = \{40.0,59.77,80.0\}$. The conservation of $I_t$ ensures that the trajectories must follow tracks where $I_{A+B} = I_t-I_C$. These are plotted as the dashed green lines in the upper left panel. The value of $I_t$ sets the intercept of these lines with the $I_C$ axis and as we increase $I_t$ they are translated upwards. For $I_t=40$ the green line never intersects the red upper branch and so the phase portrait has no fixed points. When $I_t$ is set such that the green line just touches the upper branch this results in a saddle node bifurcation, spawning two resonant centres. By combining the conservation of action constraint with the upper branch equation, we find that this bifurcation point occurs at
\begin{equation}
    \label{nonlinear_eq:upper_saddle_node}
    I_{c,saddle} = \left[ \frac{12}{c(\gamma)2^{-\gamma/2}J_{11}\gamma}\right]^{-\frac{2}{2+\gamma}}.
\end{equation}
Increasing $I_t$ beyond this shows that the two fixed points diverge as the green dashed line intersects the upper branch in two locations. The point to the left of the saddle-node is an unstable saddle whilst the point to the right is a stable centre. This elucidates the stability structure discussed previously for the upper branch. Indeed, inputting $\gamma=5/3$ and $J_{11}=100$ into equation \eqref{nonlinear_eq:upper_saddle_node} yields $I_{c,saddle} = 9.05$, which agrees with the critical value of $I_C$ separating the stable and unstable regime as seen in Fig.~\ref{fig:nonlinear:upper_branch_stability}.

An equivalent analysis can be performed for the lower branch. Now the full set of action-angle equations are reduced by restricting our attention to the case $I_A=I_B$ and $\theta_B = \theta_A+\pi$ for which we only permit out of phase tilt and shear motions. This is the correct simplification since we found it is the out of phase tilt and shear mode which is susceptible to the parametric instability. We again reduce the dimensionality of the original system and perform the same canonical transformation as before. This is then described by the Hamiltonian 
\begin{align}
    \label{nonlinear_eq:hamiltonian_reduced_lower}
   \mathcal{H}  &= -\frac{\sqrt{2}}{J_{11}}P_1\sqrt{P_2-P_1}\left[1+\cos{(2\phi_1)}\right]+\Delta(P_2-P_1)  \nonumber\\ 
        &= -\frac{\sqrt{2}}{J_{11}}I_{A+B}\sqrt{I_t-I_{A+B}}\left[1+\cos{(2\phi_1)}\right]+\Delta(I_t-I_{A+B}).
\end{align}
\begin{figure*}
    \centering
    \includegraphics[width = 0.95\textwidth]{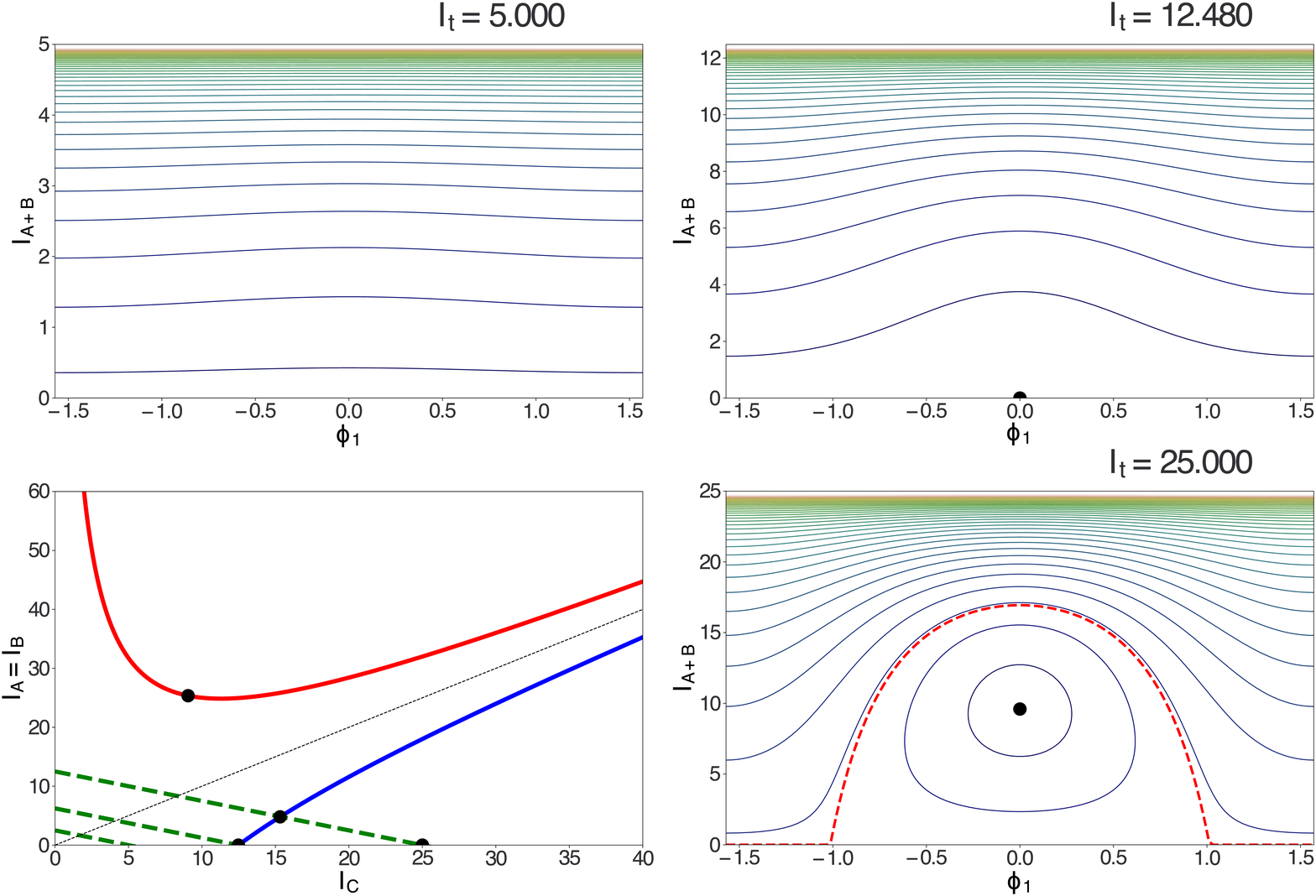}
    \caption{\textit{Top Left}: The resonant branch structure as per Fig.~\ref{fig:resonant_branches} is overlaid with green dashed lines corresponding to $I_{A+B}=I_t-I_C$ for three choices of $I_t = \{5.0,12.48,25.0\}$. Positions where the green dashed line intersects the lower branch result in fixed points, denoted by black dots. The remaining panels show contours of the Hamiltonian \eqref{nonlinear_eq:hamiltonian_reduced_lower} for the different values of $I_t$. We see that as $I_t$ is increased and the green line crosses the parametric instability threshold for the vertical breathing mode, a pitchfork bifurcation spawns two saddle-points and a centre, plotted as black dots in the bottom panels.}
    \label{fig:nonlinear:reduced_lower_portrait}
\end{figure*}
Again we can visualise this for slices through constant $I_t$ as shown in  Fig.~\ref{fig:nonlinear:reduced_lower_portrait}. The three choices of $I_t = \{5.0,12.48,25.0\}$ correspond to the three green dashed lines in the upper left panel, along which $I_A$, $I_B$ and $I_C$ are constrained to move. When $I_t=1.0$ the green line intersects the $I_C$ axis before the onset of parametric instability and we see that the purely vertical oscillator is stable. However when $I_t = 12.5$ the system crosses the bifurcation point defined by equation \eqref{nonlinear_eq:parametric_bifurcation_point}. Beyond this, we see the formation of unstable saddle-points along the $I_{A+B}=0$ axis which once again emphasises the instability of purely vertical breathing modes here. The intercept of the green dashed line with the lower branch yields the stable centre for the anti-phased mixed mode. Of course this agrees with the stable behaviour for the lower branch, found earlier using linear perturbation techniques.

\subsection{Interpreting these solutions}

Having developed a thorough understanding of our resonant equilibria it is important to relate these back to their physical interpretation. Returning to the more intuitive Jacobian coordinates, these stable action branches correspond to constant amplitude oscillatory solutions for the variables $J_{13}$, $J_{31}$ and $J_{33}$. The process of mapping from bounce to bounce of the non-linear vertical oscillator effectively removes the harmonic motion in between. In this sense, our technique effectively captures the slow timescale associated with amplitude and phase evolution.

Both resonant branches predict a highly non-linear family of bouncing modes with excited shearing and warp. Since $I_A=I_B$, the $J_{13}$ and $J_{31}$ oscillators are excited with equal amplitude, akin to the equipartition seen in the linear modes for which $\kappa=\nu$. Whilst the resonant angles are constant, $\theta_A$, $\theta_B$ and $\theta_C$ each advance at a steady rate,
\begin{equation}
    \label{nonlinear_eq:precession_rate}
    \frac{d\theta_{A,B,C}}{dn} = \pm \frac{2\sqrt{2 I_C}}{J_{11}},
\end{equation}
where the plus and minus signs correspond to the upper and lower branches respectively. The progression of tilt, shear and bouncing phase results in a precession of the modes when viewed from a global reference frame. To see this, consider a global ring with azimuthal variation in tilting and thickness. If this torus is fixed in space, an orbiting fluid parcel would see the periodic structure pass by at the orbital frequency with a fixed phase set by the azimuthal origin. However, if the structure rotates and the azimuthal origin evolves, the orbiting observer would see this time dependent phase manifest as a modification to the periodic frequency. This precessional frequency is then defined by
\begin{equation}
    \label{nonlinear_eq:resonant_precession_frequency}
    \omega_p \equiv -\frac{d\theta}{dt} = -\frac{d\theta/dn}{\pi-d \theta /dn}.
\end{equation}
For the upper branch we have retrograde precession $\omega_p<0$ and for the lower branch prograde precession $\omega_p>0$. Within the local model these predict periodic solutions with angular frequency $\omega = \Omega-\omega_p$, so when the global torus rotates with the orbit, the local frequency decreases. Meanwhile if the ring rotates against the orbit, the local frequency increases. \cite{Ogilvie2013} previously showed that discs with a fixed global warping geometry permit periodic solutions provided the epicyclic and vertical frequencies are sufficiently detuned or a viscosity is introduced to temper the resonant flows. Here however, we see the Keplerian resonance drives a precession of the ring which acts as an effective detuning from the orbital frequency. 
\section{Numerical verification}
\label{section:shooting_method}

The smooth modulation theory developed in section \ref{section:smooth_modulation} and the bouncing theory developed in sections \ref{section:bouncing_regime} and \ref{section:resonant_centres} may now be tested by returning to our full equation set \eqref{nonlinear_eq: J_11_ode} -- \eqref{nonlinear_eq: J_33_ode} and numerically finding the periodic solutions. We select the same parameters as described in the setup of section \ref{subsection:tilting_setup} which we will now reiterate. Of course we are examining the resonant case with $\kappa=\nu=\Omega$ and adopt units so $\Omega=L=1$. We take $\gamma=5/3$ and choose the characteristic temperature and $C_z$ circulation constant so that the equilibrium ring has $J_{11}=100$ and $J_{33}=1$, corresponding to an aspect ratio of $\epsilon=0.01$.

We proceed with the same shooting scheme previously used to identify the periodic solutions for the forced vertical oscillator in section \ref{subsubsection:forced_vertical_shoot}. However, as our analysis has shown, the feedback of the vertical oscillator onto the warp results in a phase modulation of the tilt and shear oscillations. These may be interpreted as precessing modes with a period which now deviates from the orbital timescale. Thus our shooting code is generalised to incorporate the period as a parameter which should also be determined. Furthermore, our theory predicts that the periodic solutions correspond to the nonlinear extension of bending waves for which the tilt and shear are in equipartition with phase relationship $0$ or $\pi$. We use this to inform our initial guesses in the shooting method.

\begin{figure*}
    \centering
    \includegraphics[width=0.95\textwidth]{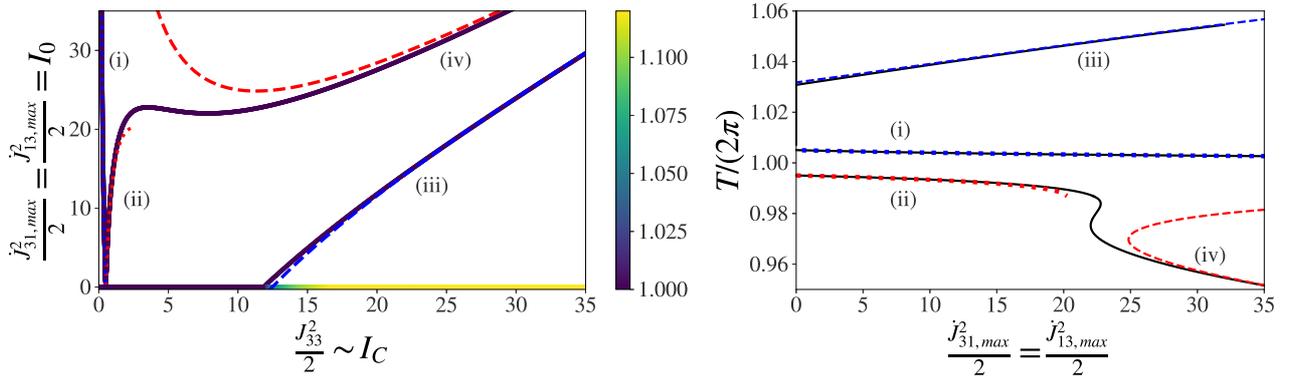}
    \caption{
    \textit{Left panel}: Branches of periodic solutions are plotted as thick lines, coloured according to the maximum eigenvalue of the monodromy matrix associated with each periodic solution. The x-axis plots the initial value of $J_{33}^2/2$ which is a measure of the vertical action $I_C$. The y-axis plots the maximum value of $\dot{J}_{13}^2/2 = \dot{J}_{31}^2/2$ which measures the warping action $I_0$. Four branches of qualitatively different solutions are labelled (i)--(iv). Branches (i) and (ii) represent the smoothly modulated anti-phased and in-phase tilt and shear solutions respectively, stemming from the equilibrium at $J_{33}=1$. The analytical predictions are over-plotted as dotted blue and red lines. Branches (iii) and (iv) then represent the extreme bouncing regime. The over-plotted blue and red dashed lines show the correspondence with theory encapsulated in the branch equations \eqref{nonlinear_eq:resonant_centres}. \textit{Right panel}: The converged period normalised against the orbital value is plotted as a black line against the $I_0$ warping action. The qualitative regimes (i)--(iv) are identified corresponding to the different solution branches in the left panel. The dotted red and blue lines over-plotting branches (i) and (ii) show the predicted period according to our smooth modulation theory. Similarly the dashed red and blue lines overlying branches (iii) and (iv) plot the period predicted from our impulsive bouncing theory. Note the in-phase branches fall below the orbital period whilst the anti-phased solutions are longer than the orbital period.
    }
    \label{fig:nonlinear:shooting_branches}
\end{figure*}

We converge to the periodic branch structure which is plotted in the left panel of Fig.~\ref{fig:nonlinear:shooting_branches}. The solid lines mark the periodic solutions found, whilst the dashed lines correspond to the solution branches predicted from our theory. The shooting method solutions are identified in terms of the Lagrangian variables, $J_{ij}$, which are then approximately converted into action variables by identifying the maximum values of $\frac{1}{2}\dot{J}_{13}^2 = \frac{1}{2}\dot{J}_{31}^2$ as the warping action $I_0=I_A=I_B$ which is plotted along the y-axis. Then the value of $J_{33}$ is extracted at times for which the forcing product $J_{13}J_{31}=0$, such that $\frac{1}{2} J_{33}^2$ is our proxy for the vertical action variable $I_{C}$ which is plotted along the x-axis. For each identified solution we perform a Floquet stability analysis, as per the method described in section \ref{subsubsection:forced_vertical_shoot}. The maximum eigenvalue from the computed monodromy matrix determines the colour along the branches, with values greater than 1 (departing from purple) indicating instability. In the right hand panel we plot the period of these solutions against the warping amplitude as the solid black lines. Again these are compared with the analytical theory predictions which are plotted as dashed and dotted lines.

\begin{figure*}
    \centering
    \includegraphics[width=0.89\textwidth]{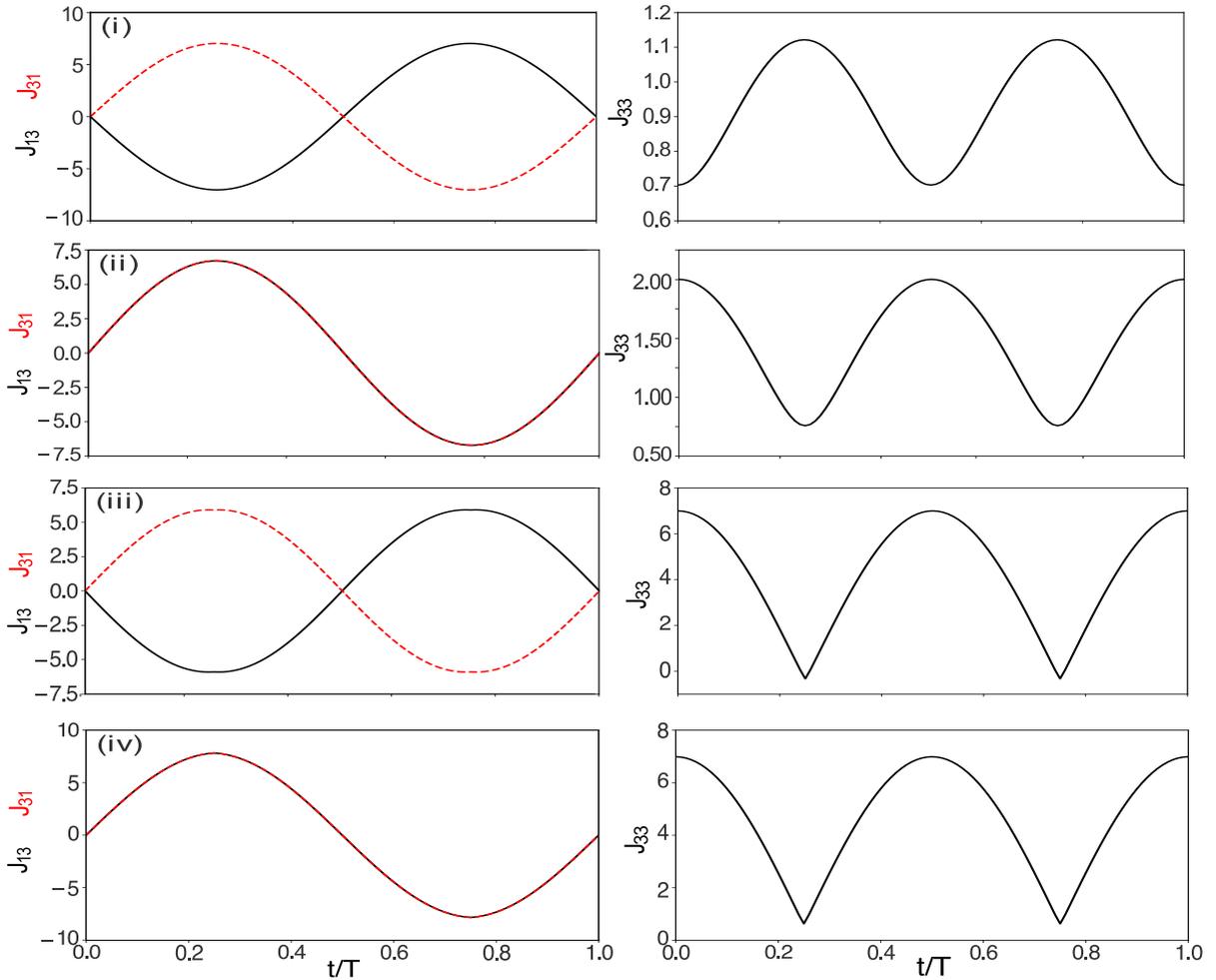}
    \caption{Example periodic solutions from the four key branches identified in Fig.~\ref{fig:nonlinear:shooting_branches} are labelled (i)--(iv). The left columns plot the $J_{13}$ shear (black) and $J_{31}$ tilt (red) variables across one period. The right hand panels then plot the corresponding evolution of $J_{33}$.
    }
    \label{fig:nonlinear:periodic_solution_grid}
\end{figure*}

We number the qualitatively distinct branches (i)--(iv), and show typical solutions for each regime in the rows of Fig.~\ref{fig:nonlinear:periodic_solution_grid}. In branches (i) and (ii) we see the anti-phased and in-phase smooth nonlinear branches respectively. These stem from the equilibrium configuration for which there is no tilt or shear and a constant value of $J_{33}=1$ and $J_{11}=100$ such that the aspect ratio of the thin base state is $\epsilon=0.01$. As we expect from the continuation of the averaged Lagrangian for $X<0$ in section \ref{subsection:modulation_precessing_solutions}, the solutions for the anti-phased tilt and shear may be continued indefinitely to large warp amplitudes. Indeed, our smooth modulation theory agrees very well as indicated by the over-plotted blue dashed line. This plots the value $J_{33}$ inherited from the periodic solutions found for equation \eqref{eq:weaklynonlinear:vertical_oscillator} at times for which the the forcing product $J_{13,0}J_{31,0} = 0$ (setting a consistent phase relationship with the warp as compared with the choice described above). The right panel of Fig.~\ref{fig:nonlinear:shooting_branches} shows that the period for this branch is slightly greater than the orbital period and agrees very well with the precessional frequency offset as deduced from the gradient of the average Lagrangian, as per equation \eqref{eq:smooth_modulation:period}. Branch (ii) meanwhile shows a more interesting behaviour. The red dashed line from the modulation theory agrees very well with the identified periodic structures for low to intermediate warp amplitudes. There is also good agreement for the predicted period within this range, which is slightly less than the orbital period as expected for the extension of the in-phase bending modes. However, the red dashed line eventually terminates at the saddle-node bifurcation, as seen for the computed average Lagrangian at some critical in-phase forcing $X>0$ -- see Fig.~\ref{fig:avg_lag}. Beyond this point the modulation theory breaks down and we expect some different behaviour to arise. 

Here, the periodic solutions begin to deviate from our modulation theory and bend round onto branch (iv). Now the vertical oscillator action begins to grow rapidly as it enters into the extreme bouncing regime. The red dashed line, showing the predicted in-phase bouncing centres as described by equation \eqref{nonlinear_eq:resonant_centres}, converges to the periodic solutions as the bounce amplitude increases. Note, the periodic solution space identified avoids the unstable portion of the upper bouncing branch since the vertical action is in fact too low here and the bouncing approximations break down. Instead there is a smooth transition connecting onto the modulation theory. The period of these in-phase bouncing solutions also shows a dramatic change in behaviour as the retrograde detuning from the orbital rate becomes more pronounced. At large warp amplitudes (and hence bouncing amplitudes), the analytical period predictions deduced from equations \eqref{nonlinear_eq:precession_rate} and \eqref{nonlinear_eq:resonant_precession_frequency} agree well.

Along the $I_C$ axis of Fig.~\ref{fig:nonlinear:shooting_branches} we see the non-linear vertical mode with no tilt and shear activation. As discussed in section \ref{subsection:branch_structure}, this undergoes parametric instability and spawns the lower anti-phased bouncing branch as labelled by (iii). Beyond this point the departure from purple colouration emphasises the instability of the pure bouncing mode with no warp activation. The lower bouncing branch incurs both growing tilt/shear and extreme bouncing motions as predicted from the lower branch of equation \eqref{nonlinear_eq:resonant_centres}. This analytical result is over-plotted as a dashed blue line which agrees remarkably well and nicely intersects the parametric instability threshold along the x-axis. This correspondence with theory is further confirmed in the period plot where there is almost perfect overlap between the dashed blue line and black line in branch (iii). We see that the bouncing solution incurs a large prograde departure from the orbital frequency as $T$ becomes longer for larger warp amplitudes. 
%
\section{Discussion}
\label{section:discussion}

These numerical results confirm our smooth modulation theory and the connection to the predicted bouncing regime. In both cases the key effect is the feedback of the vertical oscillator onto the warp which has not been taken into account in previous work. Here we see that the period of the solutions deviates from the orbital value in order to circumvent the Keplerian resonance for which $\kappa=\nu=\Omega$.

The periodic solution branches found using our local model may be reinterpreted as large-scale precessing structures when viewed from a non-rotating, global reference frame. In Paper I we saw that by Doppler shifting the linear tilting modes of our ring model into the non-rotating frame, they may be interpreted as global bending waves. Essentially the orbital time within the local model can be mapped onto the azimuthal coordinate as the shearing box performs its orbit. The ring evolution over the orbital timescale simply corresponds to the azimuthal variation in the geometry of the disc as elucidated in section \ref{section:ring_model:interpretation}.  

We might then think of our tilting ring as a model which approximately zooms in on a local patch of a globally warped disc. Thus we expect the qualitative solution families found in this paper to be applicable to a radially extended, globally warped Keplerian disc. The linear tilting modes extend into branches (i) and (ii) where the smooth modulation theory applies. The anti-phased solutions have prograde precession whilst the in-phase solutions exhibit a retrograde precession. We have focused on finding special periodic solutions for which the amplitude of the tilt and shear are constant. However, we might speculate that the combination of general tilt and shear initialisations, as plotted in the middle panel of Fig.~\ref{fig:warp_amplitude} and the upper four panels of Fig.~\ref{fig:ring_solver_run} for example, might be some modified superposition of these  nonlinear precessing modes. We see that the retrograde precession dominates over the prograde precession as the warp amplitude increases and branch (ii) bends away from the orbital period in the right panel of Fig.~\ref{fig:nonlinear:shooting_branches}. Hence we can expect a typical retrograde bias for warped structures. 

Crucially we found that this behaviour breaks down as the tilt and shear grow to sufficient amplitudes. For $\gamma=5/3$ we found that the smooth modulation theory breaks down for $Z_{1,c} = 0.4$, where the solutions for the forced vertical oscillator terminate in a saddle node bifurcation. More generally, $Z_1$ can can be connected with the global warping amplitude by the following scaling argument. Consider the radial tilting of the reference midplane line $z_0=0$ and the shearing of the vertical axis $x_0=0$. If the vertical and horizontal displacements from equilibrium are denoted by $\xi_z$ and $\xi_x$ respectively, the associated gradients are $\partial\xi_z/\partial x \sim J_{31}/J_{11}=\psi$ (where $\psi$ is the warp amplitude) and $\partial\xi_x/\partial z \sim J_{13}/J_{33}$. For a Keplerian bending wave, equipartition of tilt and shear energy demands that the displacements are of the same order. Identifying the typical warp length scale $\lambda$ as the width of our ring and taking the scale height $H$, the characteristic tilt and shear displacements balance provided $\partial\xi_x/\partial z \sim (\psi\lambda)/H$. Noting the relation $Z_2 \sim J_{13} J_{31} \sim J_{11} J_{33} Z_1$ as described by equation \eqref{nonlinear_eq:Z_2_Z_1} we see that 
\begin{equation}
    Z_1 \sim \frac{J_{13}J_{31}}{J_{11}J_{33}} \sim \frac{\partial \xi_z}{\partial x}\frac{\partial \xi_x}{\partial z} \sim \psi^2 \frac{\lambda}{H},
\end{equation}
so the critical warp amplitude scales as $\psi_c \sim \sqrt{H/\lambda}$, i.e. as $\sqrt{H/r}$ in the case of a global warp ($\lambda \sim r$). Whilst we have used $\epsilon=0.01$ throughout the course of this paper to emphasise that the warping length scale is much longer than the disc scale-height, our results also extend through to thicker discs with $\epsilon=0.1$ where we have verified the critical warp scaling law above. Beyond this value of warp, we would expect extreme vertical bouncing motions to be activated in the warped disc. This would correspond to the transition towards branch (iv), where the global warped geometry indicated by the oscillating $J_{31}$ component is now accompanied by extreme compression of $J_{33}$ twice per orbit, as seen in Fig.~\ref{fig:nonlinear:periodic_solution_grid}. The disc would present locations which are extremely thin, whilst other regions are vertically extended. This may lead to observational signatures sensitive to enhanced density. Furthermore, puffed up regions or sufficient warp amplitudes may obscure light from a central source and cast shadows as found in the various observations discussed in section \ref{subsection:intro:astrophysical_motivation}. 

The solution families predicted here are found using ideal hydrodynamics where we have no dissipation, despite the extreme compressive behaviour. However, by incorporating some viscosity prescription and a more general energy equation we might expect that the compressive motions would lead to a significant damping of the warp. Indeed, similar `nozzle-like' compressive structures occur in eccentric disc models of tidal disruption events (TDEs) wherein bouncing modes are forced periodically as gravity is enhanced at pericenter \citep{Ogilvie2014,Lynch2020}. These motions may release significant amounts of energy as the gas is compressed at closest approach \citep{ZanazziOgilvie2020,Ryu2021}. Furthermore, global warped disc simulations performed by \cite{Sorathia2013} exhibit an enhanced damping of the warp. This is not explained in their paper but might be attributed to the conversion of warp action to extreme vertical motions, via the nonlinear mode coupling, which is then damped due to the bulk artificial viscosity. Future numerical work should examine if these extreme phenomena are in fact present in the simulations and then establish observational consequences.

In fact the prediction of a critical warp amplitude in our work is reminiscent of the recent quest to understand ring breaking phenomena which are believed to occur in sufficiently warped discs \citep[e.g.][]{Nixon2012,Dogan2018}. The interplay and connection between our critical warp with this previous work is unclear and merits future investigation. Indeed, a variety of other effects might modify our solution families, including the parametric instability proposed by \cite{Gammie2000}. This has been shown to be active in global disc simulations by \cite{Deng2021} and may present an enhanced turbulent viscosity affecting the evolution of our periodic modes. In the future, we propose setting up numerical simulations which target the internal flow structure of warped discs and test how robust they are in the presence of more general physics.

\section{Conclusions}
\label{section:conclusions}

In this paper we have performed an extensive nonlinear analysis of the local ring model equations derived in Paper I and uncovered two distinct regimes relevant to the nonlinear dynamics of warped Keplerian discs. We find the extension of the linear bending modes at larger warp amplitudes is well described using an asymptotic averaged Lagrangian theory whereby the amplitude and phase of the warp smoothly vary over a long timescale. However, beyond some critical warp (which scales as the aspect ratio of the ring or disc), the in-phase product of tilt and shear motions resonantly force the vertical oscillation to large amplitudes. The disc becomes extremely compressed and feeds back impulsively onto the warp. We have identified periodic solutions using a variety of careful approximations which have then been confirmed within the full equation set. These local modes map onto globally precessing warped structures with compressions and expansions twice per orbit. These regions could manifest observationally as regions of enhanced emission or by casting shadows to outer regions of the disc. Although we have analytically extracted special solutions, we expect these compressive motions to be present in more general setups, as evidenced in our motivating numerical experiments. This may have profound consequences for the evolution of warped discs as such compressions might lead to an enhanced dissipation of energy and warp in Keplerian systems. This demands attention in future numerical simulations, with detailed analysis of the flow structure as the warp amplitude is varied. 

\section*{Acknowledgements}
The authors would like to thank the anonymous reviewer for their helpful comments and suggestions. This research was supported by an STFC studentship and STFC grants ST/P000673/1 and ST/T00049X/1.

\section*{Data Availability}
 
Data used in this paper is available from the authors upon reasonable request.




\bibliographystyle{mnras}
\bibliography{mybib} 



\appendix


\bsp	
\label{lastpage}
\end{document}